\documentclass[a4paper,12pt]{article}
\usepackage[dvips]{epsfig}
\usepackage{rotating}

\newcommand{\Zbb}   {{\mathrm Z}\to {\mathrm b}\bar{\mathrm b}}
\newcommand{\bD}    {{\mathrm b}\to {\mathrm D}}

\newcommand{\cl}    {{\mathrm c}\to {\mathrm {X}}\ell \nu} 
\newcommand{\bl}    {{\mathrm b}\to {\mathrm {X}}\ell \nu}
\newcommand{\bul}   {{\mathrm b}\to {\mathrm {X_u}}\ell}
\newcommand{\bcl}   {{\mathrm b}\to {\mathrm c}\to {\mathrm X} \ell \nu}
\newcommand{\bcbl}  {{\mathrm b}\to {\mathrm W}\to  \bar{\mathrm c}\to \ell}

\newcommand{\Bs}    {{\mathrm{B}}^0_{\mathrm s}}
\newcommand{\Bd}    {{\mathrm{B}}^0_{\mathrm d}}
\newcommand{\Dstar} {{\mathrm{D}}^\star}
\newcommand{\Dand}  {{\mathrm{D}}^{(\star)}}
\newcommand{\Dstst} {{\mathrm{D}}^{\star\star}}
\newcommand{\Done}  {{\mathrm{D}}_1}
\newcommand{\Dtwo}  {{\mathrm{D}}^\star_2}
\newcommand{\BRcl}  {{\mathrm{BR}}({\mathrm c}\to \ell)}
\newcommand{\BRbl}  {{\mathrm{BR}}({\mathrm b}\to {\mathrm {X}} \ell \nu)}
\newcommand{\BRbul} {{\mathrm{BR}}({\mathrm b}\to {\mathrm {X_u}} \ell)}
\newcommand{\BRbcl} {{\mathrm{BR}}({\mathrm b}\to {\mathrm c}\to {\mathrm X} \ell \nu)}
\newcommand{\BRbcbl}{{\mathrm{BR}}({\mathrm b}\to {\mathrm W}\to  \bar{\mathrm c}\to \ell)}
\newcommand{\chibar}{\bar \chi}
\newcommand{\Vcb}   {|V_{\mathrm{cb}}|}
\newcommand{\ups}   { \Upsilon\mathrm{(4S)}}
\newcommand{\BB}    {\mathcal {B}}
\newcommand{\JJ}    {\mathcal {J}}
\newcommand{\LL}    {\mathcal {P}}
\newcommand{\thr}   { |{\cos\theta _{\mathrm{thrust}}}|}
\newcommand{\btag}  {\mathcal {B}_{\mathrm{tag}}}

\newcommand{\pt}    {p_\perp}
\newcommand{\ptbeam}{p_{\mathrm{T}}}

\newcommand{\gevc}  {{\mathrm{GeV}} \! /c }

\newcommand{\mevc}  {{\mathrm{MeV}} \! /c }
\newcommand{\Pb}    {P_{{\mathrm{b}}\bar{\mathrm{b}}}}
\newcommand{\Pbdt}  {P_{{\mathrm{b}}\bar{\mathrm{b}}}^{\mathrm{DT}}}
\newcommand{\Pbm}   {P_{\mathrm{b}} ^{\mathrm{DT}}  }
\newcommand{\Pbp}   {P_{\bar{\mathrm{b}}} ^{\mathrm{DT}} }
\newcommand{\Pbl}   {P_{{\mathrm{b}}\bar{\mathrm{b}}}^{\mathcal {P}}}
\newcommand{\Pbj}   {P_{{\mathrm{b}}\bar{\mathrm{b}}}^{\mathcal {J}}}
\newcommand{\Pcl}   {P_{{\mathrm{c}}\bar{\mathrm{c}}}^{\mathcal {P}}}
\newcommand{\Pcj}   {P_{{\mathrm{c}}\bar{\mathrm{c}}}^{\mathcal {J}}}
\newcommand{\KS}    {{\mathrm{K}}^0_{\mathrm{S}}}
\newcommand{\BRbclh}{{\left. \BRbcl \right|}_{\mathrm h}}
\newcommand{\BRbcll}{{\left. \BRbcl \right|}_\ell}
\newcommand{\fhemi} {F_{\mathrm{hemi}}}
\newcommand{\fhemib}{f_{\mathrm{hemi}}^{\mathrm{b}}}
\newcommand{\fhemic}{f_{\mathrm{hemi}}^{\mathrm{c}}}
\newcommand{\fhemix}{f_{\mathrm{hemi}}^{\mathrm{uds}}}
\newcommand{\fevt}  {F_{\mathrm{evt}}}
\newcommand{\fevtb} {f_{\mathrm{evt}}^{\mathrm{b}}}
\newcommand{\fevtc} {f_{\mathrm{evt}}^{\mathrm{c}}}
\newcommand{\fevtx} {f_{\mathrm{evt}}^{\mathrm{uds}}}
\newcommand{\Foc}   {F^{\mathrm{oc}}}
\newcommand{\Fpp}   {F^{\mathrm{++}}}
\newcommand{\Fmm}   {F^{\mathrm{--}}}
\newcommand{\Fppb}  {F^{\mathrm{++}}_{\mathrm{b}}}
\newcommand{\Fmmb}  {F^{\mathrm{--}}_{\mathrm{b}}}
\newcommand{\Focb}  {F^{\mathrm{oc}}_{\mathrm{b}}}
\newcommand{\Focc}  {F^{\mathrm{oc}}_{\mathrm{c}}}
\newcommand{\Focx}  {F^{\mathrm{oc}}_{\mathrm{uds}}}
\newcommand{\Fppc}  {F^{\mathrm{++}}_{\mathrm{c}}}
\newcommand{\Fppx}  {F^{\mathrm{++}}_{\mathrm{uds}}}
\newcommand{\Fmmc}  {F^{\mathrm{--}}_{\mathrm{c}}}
\newcommand{\Fmmx}  {F^{\mathrm{--}}_{\mathrm{uds}}}
\newcommand{\Rb}    {R_{\mathrm b}}
\newcommand{\Rc}    {R_{\mathrm c}}

\newcommand{\bb}    {{\mathrm b}\bar{\mathrm b}}
\newcommand{\cc}    {{\mathrm c}\bar{\mathrm c}}
\newcommand{\gbb}   {N({\mathrm g}\to{\mathrm b}\bar{\mathrm b})}
\newcommand{\gcc}   {N({\mathrm g}\to{\mathrm c}\bar{\mathrm c})}
\newcommand{\gsbb}  {N({\mathrm g}\to{\mathrm b}\bar{\mathrm b})}
\newcommand{\gscc}  {N({\mathrm g}\to{\mathrm c}\bar{\mathrm c})}
\newcommand{\Jpsi}  {{\mathrm J}\! / \! \psi}
\newcommand{\BRJpsill} {{\mathrm{BR}}({\mathrm{b\to J}}\! / \! \psi\;(\psi^\prime) \to  \ell \ell)}
\newcommand{\BRbtaul}  {{\mathrm{BR}} ( {\mathrm b} \to \tau \to \ell)}

\newcommand{\eg}    {{\it{e.g.~}}}

\newcommand{\etal}{{\it{et al.}}}
\newcommand{\EPJ}   {Eur. Phys. J.}
\newcommand{\NP}    {Nucl. Phys.}
\newcommand{\NPL}   {Nucl. Phys. Lett.}
\newcommand{\NIM}   {Nucl. Instrum. Meth.}
\newcommand{\ZP}    {Z. Phys.} 
\newcommand{\PL}    {Phys. Lett.} 
\newcommand{\PRL}   {Phys. Rev. Lett.}
\newcommand{\PR}    {Phys. Rev.}

\begin{document}
\thispagestyle{empty}

\centerline{{\large EUROPEAN ORGANIZATION FOR NUCLEAR RESEARCH (CERN)}} 
\vspace{9mm}
\noindent
\begin{flushright}
\vspace{2mm}
{ CERN-EP/2001-057}\\
26 July 2001
\end{flushright}
\vskip 4 mm
\vspace{9mm}     

\begin{center}
\boldmath
{\LARGE\bf 
Inclusive semileptonic branching ratios }\\
\vskip 0.2 cm
{\LARGE\bf of b hadrons
produced in Z decays 
}\\

\vspace{1cm}
{\Large The ALEPH Collaboration}
\end{center}
\vspace{1cm}

\begin{abstract}

  \begin{sloppypar}
    
    {\noindent 
      A measurement of the inclusive semileptonic 
      branching ratios of b hadrons produced in Z decay is presented,
      using  four million hadronic events collected by the ALEPH detector
      from 1991 to 1995.
      Electrons and muons are selected  opposite to 
      b-tagged hemispheres. Two different methods are explored
      to distinguish  the contributions from direct $\bl$ and cascade $\bcl$ decays
      to the total lepton yield. One is based on the lepton transverse
      momentum spectrum,
      the other makes use
      of the correlation between the charge of the lepton
      and charge estimators built from tracks in the opposite
      hemisphere of the event. The latter method reduces the dependence
      on the modelling of semileptonic b decays.
      The results obtained by averaging the two techniques are 
      \begin{eqnarray}
        \BRbl             & = & 0.1070 \, 
        \pm 0.0010 \, _{\mathrm{stat}}\
        \pm 0.0023 \, _{\mathrm{syst}}\
        \pm 0.0026 \, _{\mathrm{model}}\ ,
        \nonumber \\
        \BRbcl            & = & 0.0818 \,
        \pm 0.0015 \, _{\mathrm{stat}} \
        \pm 0.0022 \, _{\mathrm{syst}}\
        { ^{+0.0010}  _{-0.0014}}\, _{\mathrm{model}} \ .
        \nonumber 
      \end{eqnarray}
      }
  \end{sloppypar}  
\end{abstract}
\vspace{0.8 cm}
\setlength{\textheight}{24.0cm}
\setlength{\topmargin}{-0.5cm}
\setlength{\textwidth}{15.0cm}
\setlength{\oddsidemargin}{+0.8cm}
\setlength{\topsep}{1mm}

\headsep=0.1mm

\raggedbottom
\setcounter{totalnumber}{5}
\renewcommand{\textfraction}{0.1}
\renewcommand{\floatpagefraction}{0.8}
\renewcommand{\topfraction}{0.9}
\renewcommand{\bottomfraction}{0.9}

\begin{center}
  \em  (Submitted to  European Physics Journal C) 
\end{center}

\newpage

\pagestyle{empty}
\newpage
\small
%
\newlength{\saveparskip}
\newlength{\savetextheight}
\newlength{\savetopmargin}
\newlength{\savetextwidth}
\newlength{\saveoddsidemargin}
\newlength{\savetopsep}
\setlength{\saveparskip}{\parskip}
\setlength{\savetextheight}{\textheight}
\setlength{\savetopmargin}{\topmargin}
\setlength{\savetextwidth}{\textwidth}
\setlength{\saveoddsidemargin}{\oddsidemargin}
\setlength{\savetopsep}{\topsep}
%
%
\setlength{\parskip}{0.0cm}
\setlength{\textheight}{25.0cm}
\setlength{\topmargin}{-1.5cm}
\setlength{\textwidth}{16 cm}
\setlength{\oddsidemargin}{-0.0cm}
\setlength{\topsep}{1mm}
\pretolerance=10000
\centerline{\large\bf The ALEPH Collaboration}
\footnotesize
\vspace{0.5cm}
{\raggedbottom
\begin{sloppypar}
\samepage\noindent
A.~Heister,
S.~Schael
\nopagebreak
\begin{center}
\parbox{15.5cm}{\sl\samepage
Physikalisches Institut das RWTH-Aachen, D-52056 Aachen, Germany}
\end{center}\end{sloppypar}
\vspace{2mm}
\begin{sloppypar}
\noindent
R.~Barate,
I.~De~Bonis,
D.~Decamp,
C.~Goy,
\mbox{J.-P.~Lees},
E.~Merle,
\mbox{M.-N.~Minard},
B.~Pietrzyk
\nopagebreak
\begin{center}
\parbox{15.5cm}{\sl\samepage
Laboratoire de Physique des Particules (LAPP), IN$^{2}$P$^{3}$-CNRS,
F-74019 Annecy-le-Vieux Cedex, France}
\end{center}\end{sloppypar}
\vspace{2mm}
\begin{sloppypar}
\noindent
S.~Bravo,
M.P.~Casado,
M.~Chmeissani,
J.M.~Crespo,
E.~Fernandez,
\mbox{M.~Fernandez-Bosman},
Ll.~Garrido,$^{15}$
E.~Graug\'{e}s,
M.~Martinez,
G.~Merino,
R.~Miquel,$^{27}$
Ll.M.~Mir,$^{27}$
A.~Pacheco,
H.~Ruiz
\nopagebreak
\begin{center}
\parbox{15.5cm}{\sl\samepage
Institut de F\'{i}sica d'Altes Energies, Universitat Aut\`{o}noma
de Barcelona, E-08193 Bellaterra (Barcelona), Spain$^{7}$}
\end{center}\end{sloppypar}
\vspace{2mm}
\begin{sloppypar}
\noindent
A.~Colaleo,
D.~Creanza,
M.~de~Palma,
G.~Iaselli,
G.~Maggi,
M.~Maggi,
S.~Nuzzo,
A.~Ranieri,
G.~Raso,$^{23}$
F.~Ruggieri,
G.~Selvaggi,
L.~Silvestris,
P.~Tempesta,
A.~Tricomi,$^{3}$
G.~Zito
\nopagebreak
\begin{center}
\parbox{15.5cm}{\sl\samepage
Dipartimento di Fisica, INFN Sezione di Bari, I-70126
Bari, Italy}
\end{center}\end{sloppypar}
\vspace{2mm}
\begin{sloppypar}
\noindent
X.~Huang,
J.~Lin,
Q. Ouyang,
T.~Wang,
Y.~Xie,
R.~Xu,
S.~Xue,
J.~Zhang,
L.~Zhang,
W.~Zhao
\nopagebreak
\begin{center}
\parbox{15.5cm}{\sl\samepage
Institute of High Energy Physics, Academia Sinica, Beijing, The People's
Republic of China$^{8}$}
\end{center}\end{sloppypar}
\vspace{2mm}
\begin{sloppypar}
\noindent
D.~Abbaneo,
P.~Azzurri,
G.~Boix,$^{6}$
O.~Buchm\"uller,
M.~Cattaneo,
F.~Cerutti,
B.~Clerbaux,
G.~Dissertori,
H.~Drevermann,
R.W.~Forty,
M.~Frank,
T.C.~Greening,$^{29}$
J.B.~Hansen,
J.~Harvey,
P.~Janot,
B.~Jost,
M.~Kado,
P.~Mato,
A.~Moutoussi,
F.~Ranjard,
L.~Rolandi,
D.~Schlatter,
O.~Schneider,$^{2}$
W.~Tejessy,
F.~Teubert,
E.~Tournefier,$^{25}$
J.~Ward
\nopagebreak
\begin{center}
\parbox{15.5cm}{\sl\samepage
European Laboratory for Particle Physics (CERN), CH-1211 Geneva 23,
Switzerland}
\end{center}\end{sloppypar}
\vspace{2mm}
\begin{sloppypar}
\noindent
Z.~Ajaltouni,
F.~Badaud,
A.~Falvard,$^{22}$
P.~Gay,
P.~Henrard,
J.~Jousset,
B.~Michel,
S.~Monteil,
\mbox{J-C.~Montret},
D.~Pallin,
P.~Perret,
F.~Podlyski
\nopagebreak
\begin{center}
\parbox{15.5cm}{\sl\samepage
Laboratoire de Physique Corpusculaire, Universit\'e Blaise Pascal,
IN$^{2}$P$^{3}$-CNRS, Clermont-Ferrand, F-63177 Aubi\`{e}re, France}
\end{center}\end{sloppypar}
\vspace{2mm}
\begin{sloppypar}
\noindent
J.D.~Hansen,
J.R.~Hansen,
P.H.~Hansen,
B.S.~Nilsson,
A.~W\"a\"an\"anen
\begin{center}
\parbox{15.5cm}{\sl\samepage
Niels Bohr Institute, DK-2100 Copenhagen, Denmark$^{9}$}
\end{center}\end{sloppypar}
\vspace{2mm}
\begin{sloppypar}
\noindent
A.~Kyriakis,
C.~Markou,
E.~Simopoulou,
A.~Vayaki,
K.~Zachariadou
\nopagebreak
\begin{center}
\parbox{15.5cm}{\sl\samepage
Nuclear Research Center Demokritos (NRCD), GR-15310 Attiki, Greece}
\end{center}\end{sloppypar}
\vspace{2mm}
\begin{sloppypar}
\noindent
A.~Blondel,$^{12}$
G.~Bonneaud,
\mbox{J.-C.~Brient},
A.~Roug\'{e},
M.~Rumpf,
M.~Swynghedauw,
M.~Verderi,
\linebreak
H.~Videau
\nopagebreak
\begin{center}
\parbox{15.5cm}{\sl\samepage
Laboratoire de Physique Nucl\'eaire et des Hautes Energies, Ecole
Polytechnique, IN$^{2}$P$^{3}$-CNRS, \mbox{F-91128} Palaiseau Cedex, France}
\end{center}\end{sloppypar}
\vspace{2mm}
\begin{sloppypar}
\noindent
V.~Ciulli,
E.~Focardi,
G.~Parrini
\nopagebreak
\begin{center}
\parbox{15.5cm}{\sl\samepage
Dipartimento di Fisica, Universit\`a di Firenze, INFN Sezione di Firenze,
I-50125 Firenze, Italy}
\end{center}\end{sloppypar}
\vspace{2mm}
\begin{sloppypar}
\noindent
A.~Antonelli,
M.~Antonelli,
G.~Bencivenni,
G.~Bologna,$^{4}$
F.~Bossi,
P.~Campana,
G.~Capon,
V.~Chiarella,
P.~Laurelli,
G.~Mannocchi,$^{5}$
F.~Murtas,
G.P.~Murtas,
L.~Passalacqua,
\mbox{M.~Pepe-Altarelli},$^{24}$
P.~Spagnolo
\nopagebreak
\begin{center}
\parbox{15.5cm}{\sl\samepage
Laboratori Nazionali dell'INFN (LNF-INFN), I-00044 Frascati, Italy}
\end{center}\end{sloppypar}
\vspace{2mm}
\begin{sloppypar}
\noindent
A.W. Halley,
J.G.~Lynch,
P.~Negus,
V.~O'Shea,
C.~Raine,
A.S.~Thompson
\nopagebreak
\begin{center}
\parbox{15.5cm}{\sl\samepage
Department of Physics and Astronomy, University of Glasgow, Glasgow G12
8QQ,United Kingdom$^{10}$}
\end{center}\end{sloppypar}
\vspace{2mm}
\begin{sloppypar}
\noindent
S.~Wasserbaech
\nopagebreak
\begin{center}
\parbox{15.5cm}{\sl\samepage
Department of Physics, Haverford College, Haverford, PA 19041-1392, U.S.A.}
\end{center}\end{sloppypar}
\vspace{2mm}
\begin{sloppypar}
\noindent
R.~Cavanaugh,
S.~Dhamotharan,
C.~Geweniger,
P.~Hanke,
G.~Hansper,
V.~Hepp,
E.E.~Kluge,
A.~Putzer,
J.~Sommer,
K.~Tittel,
S.~Werner,$^{19}$
M.~Wunsch$^{19}$
\nopagebreak
\begin{center}
\parbox{15.5cm}{\sl\samepage
Kirchhoff-Institut f\"ur Physik, Universit\"at Heidelberg, D-69120
Heidelberg, Germany$^{16}$}
\end{center}\end{sloppypar}
\vspace{2mm}
\begin{sloppypar}
\noindent
R.~Beuselinck,
D.M.~Binnie,
W.~Cameron,
P.J.~Dornan,
M.~Girone,$^{1}$
N.~Marinelli,
J.K.~Sedgbeer,
J.C.~Thompson$^{14}$
\nopagebreak
\begin{center}
\parbox{15.5cm}{\sl\samepage
Department of Physics, Imperial College, London SW7 2BZ,
United Kingdom$^{10}$}
\end{center}\end{sloppypar}
\vspace{2mm}
\begin{sloppypar}
\noindent
V.M.~Ghete,
P.~Girtler,
E.~Kneringer,
D.~Kuhn,
G.~Rudolph
\nopagebreak
\begin{center}
\parbox{15.5cm}{\sl\samepage
Institut f\"ur Experimentalphysik, Universit\"at Innsbruck, A-6020
Innsbruck, Austria$^{18}$}
\end{center}\end{sloppypar}
\vspace{2mm}
\begin{sloppypar}
\noindent
E.~Bouhova-Thacker,
C.K.~Bowdery,
A.J.~Finch,
F.~Foster,
G.~Hughes,
R.W.L.~Jones,$^{1}$
M.R.~Pearson,
N.A.~Robertson
\nopagebreak
\begin{center}
\parbox{15.5cm}{\sl\samepage
Department of Physics, University of Lancaster, Lancaster LA1 4YB,
United Kingdom$^{10}$}
\end{center}\end{sloppypar}
\vspace{2mm}
\begin{sloppypar}
\noindent
I.~Giehl,
K.~Jakobs,
K.~Kleinknecht,
G.~Quast,
B.~Renk,
E.~Rohne,
\mbox{H.-G.~Sander},
H.~Wachsmuth,
C.~Zeitnitz
\nopagebreak
\begin{center}
\parbox{15.5cm}{\sl\samepage
Institut f\"ur Physik, Universit\"at Mainz, D-55099 Mainz, Germany$^{16}$}
\end{center}\end{sloppypar}
\vspace{2mm}
\begin{sloppypar}
\noindent
A.~Bonissent,
J.~Carr,
P.~Coyle,
O.~Leroy,
P.~Payre,
D.~Rousseau,
M.~Talby
\nopagebreak
\begin{center}
\parbox{15.5cm}{\sl\samepage
Centre de Physique des Particules, Universit\'e de la M\'editerran\'ee,
IN$^{2}$P$^{3}$-CNRS, F-13288 Marseille, France}
\end{center}\end{sloppypar}
\vspace{2mm}
\begin{sloppypar}
\noindent
M.~Aleppo,
F.~Ragusa
\nopagebreak
\begin{center}
\parbox{15.5cm}{\sl\samepage
Dipartimento di Fisica, Universit\`a di Milano e INFN Sezione di Milano,
I-20133 Milano, Italy}
\end{center}\end{sloppypar}
\vspace{2mm}
\begin{sloppypar}
\noindent
A.~David,
H.~Dietl,
G.~Ganis,$^{26}$
K.~H\"uttmann,
G.~L\"utjens,
C.~Mannert,
W.~M\"anner,
\mbox{H.-G.~Moser},
R.~Settles,
H.~Stenzel,
W.~Wiedenmann,
G.~Wolf
\nopagebreak
\begin{center}
\parbox{15.5cm}{\sl\samepage
Max-Planck-Institut f\"ur Physik, Werner-Heisenberg-Institut,
D-80805 M\"unchen, Germany\footnotemark[16]}
\end{center}\end{sloppypar}
\vspace{2mm}
\begin{sloppypar}
\noindent
J.~Boucrot,
O.~Callot,
M.~Davier,
L.~Duflot,
\mbox{J.-F.~Grivaz},
Ph.~Heusse,
A.~Jacholkowska,$^{22}$
J.~Lefran\c{c}ois,
\mbox{J.-J.~Veillet},
I.~Videau,
C.~Yuan
\nopagebreak
\begin{center}
\parbox{15.5cm}{\sl\samepage
Laboratoire de l'Acc\'el\'erateur Lin\'eaire, Universit\'e de Paris-Sud,
IN$^{2}$P$^{3}$-CNRS, F-91898 Orsay Cedex, France}
\end{center}\end{sloppypar}
\vspace{2mm}
\begin{sloppypar}
\noindent
G.~Bagliesi,
T.~Boccali,
L.~Fo\`{a},
A.~Giammanco,
A.~Giassi,
F.~Ligabue,
A.~Messineo,
F.~Palla,
G.~Sanguinetti,
A.~Sciab\`a,
G.~Sguazzoni,
R.~Tenchini,$^{1}$
A.~Venturi,
P.G.~Verdini
\samepage
\begin{center}
\parbox{15.5cm}{\sl\samepage
Dipartimento di Fisica dell'Universit\`a, INFN Sezione di Pisa,
e Scuola Normale Superiore, I-56010 Pisa, Italy}
\end{center}\end{sloppypar}
\vspace{2mm}
\begin{sloppypar}
\noindent
G.A.~Blair,
G.~Cowan,
M.G.~Green,
T.~Medcalf,
A.~Misiejuk,
J.A.~Strong,
\mbox{P.~Teixeira-Dias},
\mbox{J.H.~von~Wimmersperg-Toeller}
\nopagebreak
\begin{center}
\parbox{15.5cm}{\sl\samepage
Department of Physics, Royal Holloway \& Bedford New College,
University of London, Egham, Surrey TW20 OEX, United Kingdom$^{10}$}
\end{center}\end{sloppypar}
\vspace{2mm}
\begin{sloppypar}
\noindent
R.W.~Clifft,
T.R.~Edgecock,
P.R.~Norton,
I.R.~Tomalin
\nopagebreak
\begin{center}
\parbox{15.5cm}{\sl\samepage
Particle Physics Dept., Rutherford Appleton Laboratory,
Chilton, Didcot, Oxon OX11 OQX, United Kingdom$^{10}$}
\end{center}\end{sloppypar}
\vspace{2mm}
\begin{sloppypar}
\noindent
\mbox{B.~Bloch-Devaux},$^{1}$
P.~Colas,
S.~Emery,
W.~Kozanecki,
E.~Lan\c{c}on,
\mbox{M.-C.~Lemaire},
E.~Locci,
P.~Perez,
J.~Rander,
\mbox{J.-F.~Renardy},
A.~Roussarie,
\mbox{J.-P.~Schuller},
J.~Schwindling,
A.~Trabelsi,$^{21}$
B.~Vallage
\nopagebreak
\begin{center}
\parbox{15.5cm}{\sl\samepage
CEA, DAPNIA/Service de Physique des Particules,
CE-Saclay, F-91191 Gif-sur-Yvette Cedex, France$^{17}$}
\end{center}\end{sloppypar}
\vspace{2mm}
\begin{sloppypar}
\noindent
N.~Konstantinidis,
A.M.~Litke,
G.~Taylor
\nopagebreak
\begin{center}
\parbox{15.5cm}{\sl\samepage
Institute for Particle Physics, University of California at
Santa Cruz, Santa Cruz, CA 95064, USA$^{13}$}
\end{center}\end{sloppypar}
\vspace{2mm}
\begin{sloppypar}
\noindent
C.N.~Booth,
S.~Cartwright,
F.~Combley,$^{4}$
M.~Lehto,
L.F.~Thompson
\nopagebreak
\begin{center}
\parbox{15.5cm}{\sl\samepage
Department of Physics, University of Sheffield, Sheffield S3 7RH,
United Kingdom$^{10}$}
\end{center}\end{sloppypar}
\vspace{2mm}
\begin{sloppypar}
\noindent
K.~Affholderbach,$^{28}$
A.~B\"ohrer,
S.~Brandt,
C.~Grupen,
A.~Ngac,
G.~Prange,
U.~Sieler
\nopagebreak
\begin{center}
\parbox{15.5cm}{\sl\samepage
Fachbereich Physik, Universit\"at Siegen, D-57068 Siegen,
 Germany$^{16}$}
\end{center}\end{sloppypar}
\vspace{2mm}
\begin{sloppypar}
\noindent
G.~Giannini
\nopagebreak
\begin{center}
\parbox{15.5cm}{\sl\samepage
Dipartimento di Fisica, Universit\`a di Trieste e INFN Sezione di Trieste,
I-34127 Trieste, Italy}
\end{center}\end{sloppypar}
\vspace{2mm}
\begin{sloppypar}
\noindent
J.~Rothberg
\nopagebreak
\begin{center}
\parbox{15.5cm}{\sl\samepage
Experimental Elementary Particle Physics, University of Washington, Seattle, 
WA 98195 U.S.A.}
\end{center}\end{sloppypar}
\vspace{2mm}
\begin{sloppypar}
\noindent
S.R.~Armstrong,
K.~Cranmer,
P.~Elmer,
D.P.S.~Ferguson,
Y.~Gao,$^{20}$
S.~Gonz\'{a}lez,
O.J.~Hayes,
H.~Hu,
S.~Jin,
J.~Kile,
P.A.~McNamara III,
J.~Nielsen,
W.~Orejudos,
Y.B.~Pan,
Y.~Saadi,
I.J.~Scott,
J.~Walsh,
Sau~Lan~Wu,
X.~Wu,
G.~Zobernig
\nopagebreak
\begin{center}
\parbox{15.5cm}{\sl\samepage
Department of Physics, University of Wisconsin, Madison, WI 53706,
USA$^{11}$}
\end{center}\end{sloppypar}
}
\footnotetext[1]{Also at CERN, 1211 Geneva 23, Switzerland.}
\footnotetext[2]{Now at Universit\'e de Lausanne, 1015 Lausanne, Switzerland.}
\footnotetext[3]{Also at Dipartimento di Fisica di Catania and INFN Sezione di
 Catania, 95129 Catania, Italy.}
\footnotetext[4]{Deceased.}
\footnotetext[5]{Also Istituto di Cosmo-Geofisica del C.N.R., Torino,
Italy.}
\footnotetext[6]{Supported by the Commission of the European Communities,
contract ERBFMBICT982894.}
\footnotetext[7]{Supported by CICYT, Spain.}
\footnotetext[8]{Supported by the National Science Foundation of China.}
\footnotetext[9]{Supported by the Danish Natural Science Research Council.}
\footnotetext[10]{Supported by the UK Particle Physics and Astronomy Research
Council.}
\footnotetext[11]{Supported by the US Department of Energy, grant
DE-FG0295-ER40896.}
\footnotetext[12]{Now at Departement de Physique Corpusculaire, Universit\'e de
Gen\`eve, 1211 Gen\`eve 4, Switzerland.}
\footnotetext[13]{Supported by the US Department of Energy,
grant DE-FG03-92ER40689.}
\footnotetext[14]{Also at Rutherford Appleton Laboratory, Chilton, Didcot, UK.}
\footnotetext[15]{Permanent address: Universitat de Barcelona, 08208 Barcelona,
Spain.}
\footnotetext[16]{Supported by the Bundesministerium f\"ur Bildung,
Wissenschaft, Forschung und Technologie, Germany.}
\footnotetext[17]{Supported by the Direction des Sciences de la
Mati\`ere, C.E.A.}
\footnotetext[18]{Supported by the Austrian Ministry for Science and Transport.}
\footnotetext[19]{Now at SAP AG, 69185 Walldorf, Germany.}
\footnotetext[20]{Also at Department of Physics, Tsinghua University, Beijing, The People's Republic of China.}
\footnotetext[21]{Now at D\'epartement de Physique, Facult\'e des Sciences de Tunis, 1060 Le Belv\'ed\`ere, Tunisia.}
\footnotetext[22]{Now at Groupe d' Astroparticules de Montpellier, Universit\'e de Montpellier II, 34095 Montpellier, France.}
\footnotetext[23]{Also at Dipartimento di Fisica e Tecnologie Relative, Universit\`a di Palermo, Palermo, Italy.}
\footnotetext[24]{Now at CERN, 1211 Geneva 23, Switzerland.}
\footnotetext[25]{Now at ISN, Institut des Sciences Nucl\'eaires, 53 Av. des Martyrs, 38026 Grenoble, France.} 
\footnotetext[26]{Now at INFN Sezione di Roma II, Dipartimento di Fisica, Universit\'a di Roma Tor Vergata, 00133 Roma, Italy.} 
\footnotetext[27]{Now at LBNL, Berkeley, CA 94720, U.S.A.}
\footnotetext[28]{Now at Skyguide, Swissair Navigation Services, Geneva, Switzerland.}
\footnotetext[29]{now at Honeywell, Phoenix AZ, U.S.A.}
%
\setlength{\parskip}{\saveparskip}
\setlength{\textheight}{\savetextheight}
\setlength{\topmargin}{\savetopmargin}
\setlength{\textwidth}{\savetextwidth}
\setlength{\oddsidemargin}{\saveoddsidemargin}
\setlength{\topsep}{\savetopsep}
\normalsize
\newpage
\pagestyle{plain}
\setcounter{page}{1}

\begin{sloppypar}

\section{Introduction}
\label{sec:intro}

Accurate knowledge of the direct inclusive semileptonic branching fraction 
of b hadrons, \mbox{$\BRbl$} ($\ell = {\mathrm{e\ or}}\ \mu$),
provides the opportunity to test and improve 
the theoretical models describing the dynamics of heavy hadrons. 
It is also an important input to a 
determination of $\Vcb$~\cite{refVcb1,refVcb2}.
Together with the cascade decay branching fraction 
$\BRbcl$, \mbox{$\BRbl$} is an important 
input parameter for many heavy flavour analyses.

Previous determinations of $\BRbl$ performed  at the Z and the $\ups$
have shown some disagreement,
with that measured at the  Z being higher  while the opposite would 
be expected from the short b baryon lifetime~\cite{pdg1998}. 
On the other hand, theoretical predictions have tended to be higher 
than experimental measurements although 
recent calculations,  including
higher order perturbative QCD corrections, give values
in better agreement~\cite{latest_teo1,latest_teo2}.

In this paper two new analyses are presented,  based on the data
collected with the ALEPH detector  from 1991 to 1995.
Two different methods are used to distinguish  the contributions from the
direct and cascade decays to the total lepton yield. 
One method has better statistical precision but a stronger 
dependence on the modelling of the semileptonic decays.
The other is designed to have minimal decay modelling dependence.
The efficiency of lepton identification is measured
from data using several control samples.
The description of the fragmentation of b quarks into b hadrons
is based on the spectrum reconstructed with the ALEPH 
data~\cite{tom} and is therefore independent of modelling assumptions.

\section{The ALEPH detector}
\label{sec.aleph}

A detailed description of the ALEPH detector can be found in~\cite{detect}
and of its performance in~\cite{aleph_perf}.
A brief overview will be given here, along with some basic 
information about lepton identification.

Charged particles are detected in the central part of the apparatus
which consists of a high resolution silicon vertex detector (VDET), 
a cylindrical drift chamber (ITC) and a large time projection
chamber (TPC).

The three tracking devices are immersed in a 1.5~T axial magnetic field
provided by a superconducting solenoid; the particle momentum 
transverse to the beam axis is measured with a resolution 
$\sigma(p_{\mathrm T})/p_{\mathrm T} = 6
\times10^{-4} p_{\mathrm T} \oplus 5\times10^{-3}$, with
$p_{\mathrm T}$ measured in $\gevc$.  
In the following, charged tracks are defined as ``good'' if a) they 
originate from within a cylinder, coaxial with the beam and centred
around the nominal interaction point, with length 20~cm  and radius 2~cm;
b) their polar angle with respect to the beam satisfies the requirement   
$|{\cos\theta}|< 0.95$; and 
c) at least four hits are reconstructed in the TPC.

Hemispheres containing  b quarks are identified with a lifetime b-tagging
algorithm~\cite{ian} which exploits the three-dimensional impact
parameter resolution of the charged tracks. For tracks with two space points 
in the VDET ({\it i.e.} $|{\cos \theta}|< 0.7$) the resolution can be parametrized
as $(25+95/p)\,\mu {\mathrm {m}}$, with $p$ measured in $\gevc$.

The TPC also provides up to 338 measurements of the specific ionization 
energy loss ($dE/dx$)
which allows electrons to be separated from other charged particles by more than
three standard deviations up to  a momentum of 8~$\gevc$.

Electrons are identified by the characteristic longitudinal and
transverse development of their associated showers in the 22 radiation
length electromagnetic calorimeter (ECAL). 
The ECAL is segmented in $0.9^\circ \times 0.9^\circ$ projective towers,
read out in three longitudinal stacks;
the energy resolution achieved for isolated electrons and photons is 
$\sigma_E/E = 0.009 + 0.18/ \sqrt{E}$, with
$E$ measured in GeV.  The $dE/dx$ information provided by the TPC enhances 
the hadron rejection power, while non-prompt electrons originating
from photon conversions in the material of the detector are rejected on the basis 
of their kinematic and geometric properties.

Muons are identified by their penetration pattern in the hadronic
calorimeter (HCAL); the additional three-dimensional coordinates
measured in two layers of external muon chambers help in resolving  
the remaining possible ambiguities.

The standard lepton identification technique is described 
in detail in~\cite{hfl}. 
For the analyses presented in this paper, 
the selection is improved  and optimized to reduce
the systematic uncertainty on the results, as discussed
in Section~\ref{sec:lep_id}.

\section{Event samples}
\label{sec:method}
The analysis is based on nearly four million hadronic Z decays
selected using charged track information~\cite{hadsel}.
This data set was reprocessed during 1998 using improved
reconstruction algorithms. The main benefits for this analysis
are related to the enhanced particle identification capabilities,
and are discussed in the description of the lepton selection
(Section~\ref{sec:lep_id}).

The statistics available for the simulation are larger than the 
data statistics by a factor of two; an additional sample of 
about five million $\Zbb$ simulated events is used for the estimate
of the correlation between the b-tagging probability and the lepton 
selection efficiency (Section~\ref{sec.hcorr}).
Each event is divided into two hemispheres by the plane
perpendicular to the thrust axis. Three samples of lepton candidates
are selected as follows.

\begin{figure}[tb!]
\begin{center}

 \mbox{\hskip -1cm \epsfig{figure=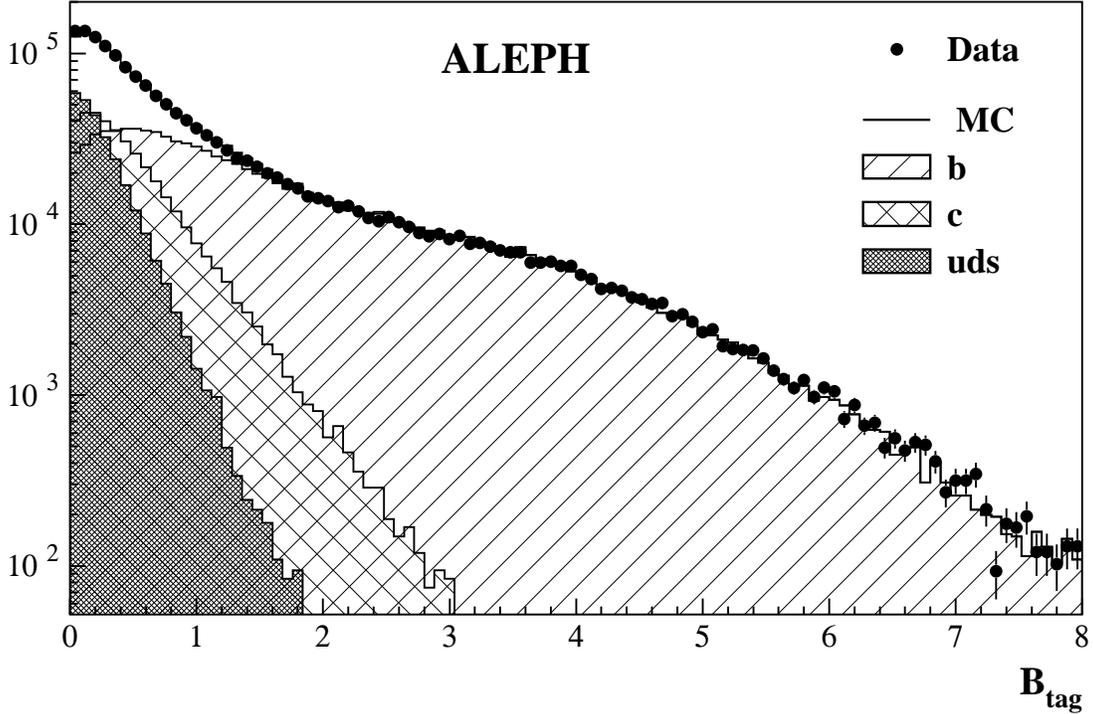,width=170mm}} 
\caption{\small Distribution of the b-tagging variable built by combining both lifetime and mass information  of tracks in each hemisphere opposite to a lepton candidate.
The error bars show the statistical error in the data only.}
\label{btag_distr}
\end{center}
\end{figure}

\paragraph{\bf\boldmath Sample $\BB$.} 
A b-tagging variable $\btag$,
based on the large mass and lifetime of b hadrons,
is built as in~\cite{ian} (Fig.~\ref{btag_distr}). 
This variable is defined using tracks contained
in one hemisphere, but here the  primary vertex is reconstructed using 
all tracks of the events, in contrast with~\cite{ian}. The impact parameter
in the simulation has been smeared in order to reproduce the
resolution measured in the data, following the same procedure
as in~\cite{ian}. The algorithm has
good performance for events well contained in the vertex detector acceptance;
events with $\thr > 0.7$ are rejected.

A cut $\btag > 2$ is applied, selecting 345,555 hemispheres in the data,
with a b purity of 97\%. The cut is about 32\% efficient on b hemispheres
within the angular region considered.

Lepton  candidates (electron or muon) are searched for
in hemispheres opposite to the selected ones.
Events in which both hemispheres are b-tagged
are used twice. The candidates are ordered according to their 
transverse momentum $\pt$, 
measured with respect to the jet axis as in~\cite{hfl}.
When more than one lepton is found in a given hemisphere,
the two leptons with highest  $\pt$ and with
opposite-charge  are used for the analysis; 
if all have the same charge the one with the highest $\pt$ is taken.
A total of 80,730 lepton candidates are finally selected.

\paragraph{\bf\boldmath Sample $\LL$.} 

\begin{figure}[tb!]
\begin{center}
 \mbox{\epsfig{figure=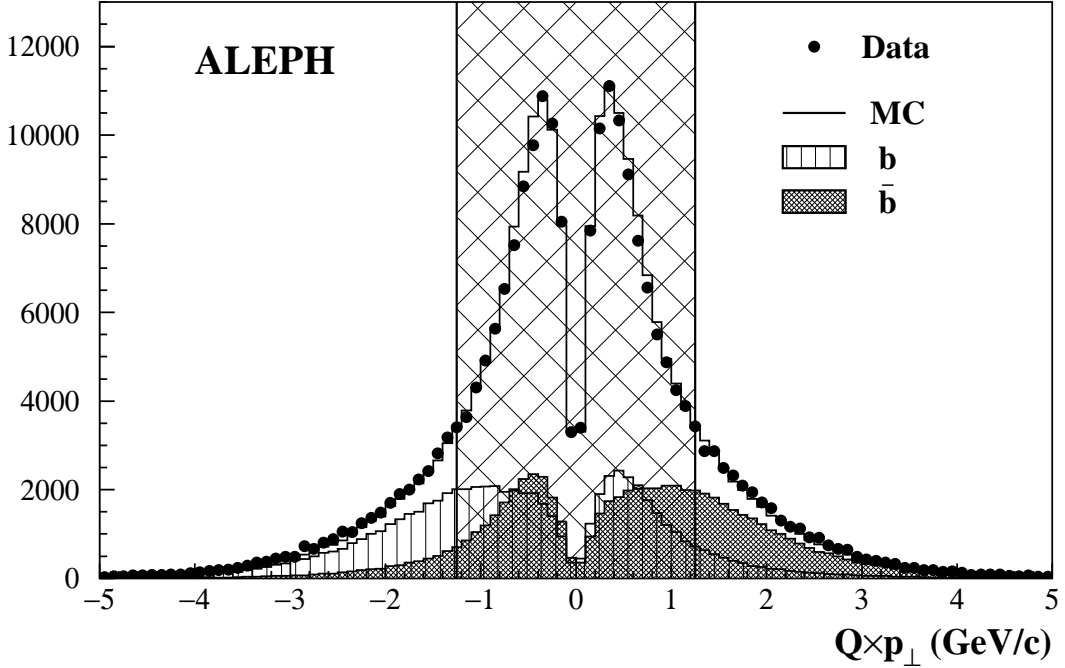,width=160mm}} 
\caption{\small The charge estimator $Q\times\pt$, based on the charge of a high $\pt$ lepton: 
data are compared with Monte Carlo simulation. 
The contributions of hemispheres containing b and $\overline{\mathrm b}$ 
quarks extracted from Monte Carlo are also shown. 
The cross-hatched area indicates the region rejected by the selection cut.}
\label{jpt_charge}
\end{center}
\end{figure}

In each event one of the two hemispheres is randomly chosen, and
a lepton candidate fulfilling the requirement 
\mbox{$\pt > 1.25\ \gevc$} is searched for.
This selects 148,001 hemispheres  in the data
with an estimated b purity of 90\% and a b efficiency of 12\%.
Since the $\pt$ cut suppresses
cascade decays, the charge $Q$ of the lepton is a good estimator of the 
charge of the  quark at production time,
leading to a  probability of tagging the charge
correctly of \mbox{$\Pbl = 0.81\pm0.01$};
this includes the dilution due to neutral B meson mixing. 
This value is measured from the data with the method described later
in Section~\ref{sec:chargetag}, and is in good agreement with the prediction
of the simulation, \mbox{$\Pbl({\mathrm{MC}}) = 0.817\pm0.004_{\mathrm {stat}}$}.
As this selection does not rely specifically on 
the vertex detector, no cut is made on $\thr$. The distribution
of $Q\times\pt$ is shown in Fig.~\ref{jpt_charge}.

Leptons are searched for in hemispheres opposite 
to the tagged ones, as in the case of the $\BB$ sample,
yielding a total of 14,509 candidates.

The {\em a priori} random choice of the hemisphere used for charge
tagging ensures that there is no double counting of lepton pairs,
and allows the measurement of $\Pbl$ from the data
(Section~\ref{sec:chargetag}).

\paragraph{\bf\boldmath Sample $\JJ$.}

Two hemisphere charge estimators are calculated:  
\begin{eqnarray*}
Q_p = \frac {\sum q \, p_\parallel^{\kappa_p}} {\sum p_\parallel^{\kappa_p}} \ ,\\
Q_s = \frac {\sum q \, s^{\kappa_s}} {\sum s^{\kappa_s}} \ ,
\end{eqnarray*}
\noindent where the sum runs over all the good charged tracks, as defined  
in Section~\ref{sec.aleph}, with momentum
in excess of $200\ \mevc$, $q$ is the charge, 
$p_\parallel$ the component of the momentum
parallel to the thrust axis, $s$ is the impact parameter significance, defined
as in~\cite{ian}, $\kappa_p = 0.5$ and $\kappa_s = 0.3$.  
Tracks with negative impact parameter are not included in the
definition of $Q_s$.

The two charge estimators are combined using weights $w$, parametrised as a function
of their magnitudes to achieve optimal performance on b hemispheres:
\[
Q_H = w(| Q_p|, | Q_s|)  \, Q_p + [1-w(| Q_p|, | Q_s|)] \, Q_s \ .
\]

\begin{figure}[tb!]
\begin{center}
 \mbox{\epsfig{figure=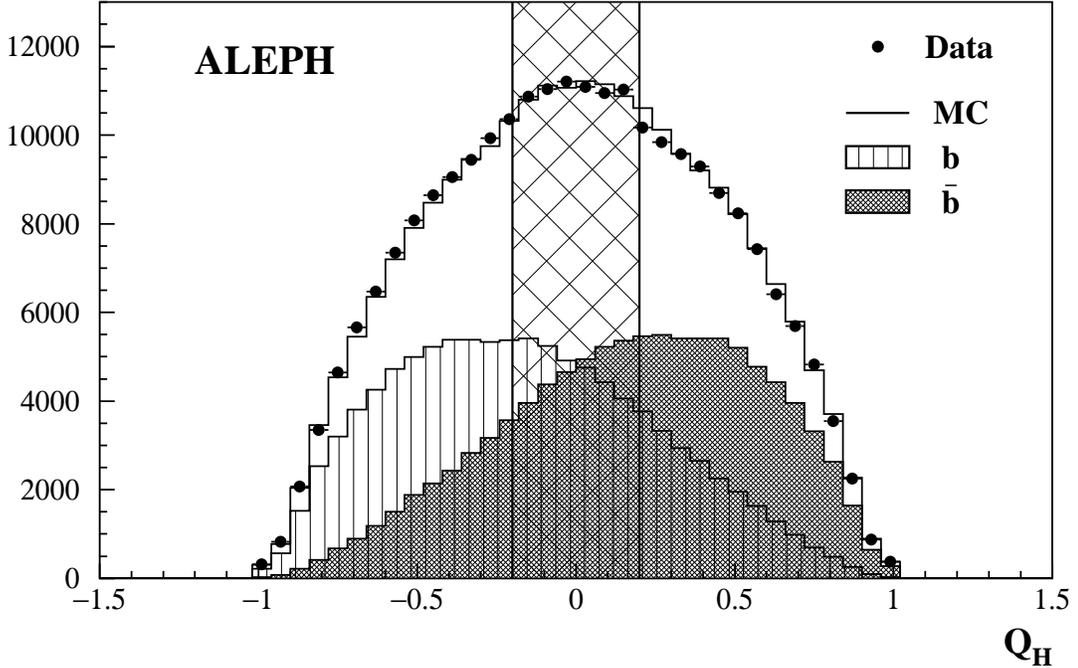,width=160mm}} 
\caption{\small Distribution of the hemisphere charge estimator $Q_H$, in data and simulation.
The contributions of hemispheres containing b and $\overline{\mathrm b}$ 
quarks extracted from Monte Carlo are also shown. 
The cross-hatched area indicates the region rejected by the selection cut.}
\label{jetq_comb}
\end{center}
\end{figure}

Hemispheres are selected if they satisfy $\btag > 1.2$, which enhances the
b content of the sample. In addition a cut $| Q_H | > 0.2$ is applied to 
ensure a good probability
that the sign of $Q_H$ is correlated with the charge of the b quark in the parent b hadron.
The distribution of the $Q_H$ variable is shown in Fig.~\ref{jetq_comb},
together with the contributions of hemispheres containing b and ${\overline{\mathrm b}}$ 
quarks.
Since the $\btag$ variable is used, events with  \mbox{$\thr > 0.7$} are not considered.
Hemispheres containing a lepton candidate with 
\mbox{$\pt > 1.25$ $\gevc$} are rejected in order to keep this sample statistically independent
of sample  $\LL$.

The procedure selects  392,523 hemispheres in the data
with an estimated b purity of 87\% and a b efficiency of 32\%.
The probability of correct b charge tagging
measured from the data is $\Pbj = 0.724 \pm 0.007$, again in good agreement with the
Monte Carlo prediction  $\Pbj({\mathrm{MC}}) = 0.721 \pm 0.001_{\mathrm {stat}}$.

The lepton yield in hemispheres opposite to the selected ones is studied 
as for the previous samples, giving a total of 61,012 candidates.

\section{Lepton identification}
\label{sec:lep_id}

A good control of the identification efficiency, as well as of the background
in the selected sample, is crucial for this analysis. 
The identification of electrons and muons follows the lines of~\cite{hfl}.
A reduced dependence on the description of the b fragmentation is achieved by
extending the acceptance to lower momentum leptons. 

The main change is, however, 
the use of a new estimator for the charged particle 
energy loss in the TPC. In the reprocessing
of the LEP I data sample, information from the pulse height 
measured by the TPC
pads has been used to build a new estimator similar 
to that from the wire measurements. 
The pad estimator is available for all tracks, 
while the wire estimator is  calculated only for tracks that have 
a minimum number of  isolated
wire signals, which leads to an average inefficiency of about 15\% 
in the hadronic environment.

The wires estimator is defined as
\[
I^{\mathrm{w}}_i = \frac{I^{\mathrm{w}}_{\mathrm{meas}} - \langle I^{\mathrm{w}} \rangle_i}
{\sigma^{\mathrm{w}}_i},
\]
\noindent where $\langle I^{\mathrm{w}} \rangle_i$ is the average energy loss
expected for particle type $i$, $\sigma^{\mathrm{w}}_i$ is the expected width,
and $I^{\mathrm{w}}_{\mathrm{meas}}$ is the measured ionization.
A similar equation defines the pad estimator $I^{\mathrm{p}}_i$.

A new energy loss estimator $I$ is built from those two. 
This estimator is just the pad estimator
for tracks that do not fulfil the requirement on the minimum 
number of isolated wire signals, otherwise
it combines wire and pad information using the B.L.U.E.
technique~\cite{blue}, taking into account the correlation of about 50\%
between the two measurements.

\subsection{Electrons}
\label{sec:el_id}

For this analysis the momentum cut is lowered to $p > 2\ \gevc$
with respect to the selection described in~\cite{hfl}.
The requirement on the minimum number of isolated TPC wire signals is dropped, 
and a cut on the new energy loss estimator  $ I_{\mathrm{e}} > -2$ is applied.
This removes the dependence of the identification efficiency upon the track
isolation, and hence the electron $\pt$. Compared to the previous selection,
the new treatment of the energy loss information provides an increase 
in efficiency
that goes from a few percent at high $\pt$ to about 30\% at low $\pt$. 
The background increases marginally.

\subsection{Muons}
\label{sec:mu_id}

The momentum cut is lowered to $p > 2.5\ \gevc$. This still ensures that all  muons
reach the muon chambers but introduces a small dependence in the identification 
efficiency on the momentum. 
A dedicated study has been performed using $\gamma\gamma \to \mu^+ \mu^-$ events 
(Section~\ref{sec:musys}).

A cut on the polar angle $| \cos \theta |<0.7$ is applied, which ensures that
the muon is within the acceptance of both vertex detector layers. At least
one VDET hit is required to be associated with the muon track, and the 
distance to the primary vertex in the plane orthogonal to the beam axis is required to satisfy 
$|d_0| < 2.5\ {\mathrm{mm}}$. This substantially reduces the contamination from
muons coming from kaon and pion decays.

In addition a cut on the energy-loss estimator $ I_\mu >-2$ is applied;
this further  reduces the background from misidentified kaons by more than a factor of two.

\section{The analysis method}
\label{sec:fit}

The lepton production rate in a sample of hemispheres with high b purity is
dominated by the $\bl$ and $\bcl$ processes. In order to
disentangle these two contributions, two different procedures
are used, one based on the lepton kinematics, the other on the
lepton charge. The two methods are described below.

\subsection{Transverse momentum analysis}

Sample $\BB$, described in Section~\ref{sec:method}, contains a large number
of nearly unbiased b decays.
The lepton rate in such a sample can be directly interpreted in terms of the
sum of the direct and cascade inclusive semileptonic b decays, weighted
by their selection efficiencies, once contributions from other physics
sources and misidentified leptons have been corrected for.

The study of the lepton rate as a function of the transverse momentum
which discriminates the two components,
allows the two branching ratios to be fitted simultaneously. 

Therefore a measurement of $\BRbl$ and $\BRbcl$ can be obtained 
from a maximum likelihood fit to the number of leptons
in each transverse momentum bin, as follows:
\begin{equation}
  {\mathcal{L}} \ = \  \prod_i \frac{e^{-\mu_i}\; \mu_i^{n_i}}{n_i!} \ ,
\label{eq:like1}
\end{equation}
\noindent where the product runs over the transverse momentum bins, $n_i$ is the number
of leptons found in the data in each bin, $\mu_i$ is the expected number of leptons, 
which depends on the two branching ratios and contains the contributions from all the
other sources of lepton candidates.

Eq.~\ref{eq:like1} can be rewritten as 
\begin{equation}
  {\cal L}  \ = \  \underbrace{ \frac{e^{-\mu}\; \mu^{N}}{N!}}_{\mathrm{counting}}
  \ \times \ \underbrace{ \prod_{j=1}^{N}
    {\cal F}(\pt^j) }_{ {\pt \ \mathrm{spectrum} }} \ ,
  \label{eq:like2}
\end{equation}
where $N$ is the total number of candidates observed in the data ($N = \sum_i n_i$), 
$\mu$ is the expected number ($\mu = \sum_i \mu_i$), and ${\cal F}(\pt)$ is the binned 
function that gives the expected shape of the distribution of the candidates as a 
function of $\pt$ (${\cal F}_i = \mu_i / \mu$). The product runs over the lepton candidates.

The expected number of candidates ($\mu$) in the selected hemispheres ($N_{\mathrm{hemi}}$)
depends on the signal branching ratios multiplied by the average selection efficiencies
$\epsilon_{{\mathrm b}\to {\mathrm {x}}}$,
and on the rates of the other sources of lepton candidates:
\[
\mu = N_{\mathrm{hemi}} \times \left[ \BRbl \epsilon_{\bl} +\BRbcl \epsilon_{\bcl} + {\mathrm{other\ sources}} \right] \ . 
\]

The expected $\pt$ shape depends on the shapes of the different components and their relative
abundances:
\[
  {\cal F}(\pt) = 
f_{\bl}  {\cal F}^{\bl}(\pt)  + 
f_{\bcl} {\cal F}^{\bcl}(\pt) + \cdots 
\]
\noindent where the shapes ${\cal F}^i(\pt)$ are normalized to unity, the $\bl$ fraction is
written as
\[
f_{\bl} = \frac{ 
\BRbl  \epsilon_{\bl} }{
\BRbl  \epsilon_{\bl}   + 
\BRbcl \epsilon_{\bcl}  + {\mathrm{other\ sources}} }\ , 
\]
and similar equations hold for all the components.

The part of the likelihood labelled as ``counting'' contains the information on the total
rate, and is therefore sensitive to the (weighted) sum of the two branching ratios. It is
affected by uncertainties on the lepton identification efficiency and background, as well
as on the rates of other processes yielding leptons. It 
has little dependence on the modelling of the $\pt$ spectrum.

The part labelled as ``$\pt$ spectrum'' is sensitive to the relative contribution of the two
signal sources, but almost insensitive to their overall normalization. 
It is heavily affected by uncertainties in the b decay modelling.

\subsection{Charge correlation analysis}

Another way of discriminating the $\bl$ and $\bcl$ components is to exploit the
different correlation of the lepton and the parent quark charges.
The second part of the likelihood in
Eq.~\ref{eq:like2} can be replaced with a term containing the fraction
of leptons that have the same or the opposite charge relative to a charge estimator
built using tracks in the opposite hemisphere (\eg jet charge). 

However, such a method would have  poor statistical power if
applied to sample $\BB$.
The charge tag samples $\LL$ and $\JJ$ are
selected for this reason, somewhat relaxing the requirement on the b purity in favour of 
higher statistics and better charge tagging.

A likelihood function is constructed based on the counting part 
from sample $\BB$ and the charge spectra of samples $\LL$ and $\JJ$ as follows:
\begin{equation}
  {\cal L}  \ =  \  \underbrace{ \frac{e^{-\mu}\; \mu^{N}}{N!}}_{\mathrm{counting\ (\BB)}}
  \   \times  \ 
  \underbrace{  {\cal F}_{\LL}^{N_{O_{\LL}}} \, {(1- {\cal F}_{\LL} )}^{N_{S_{\LL}}} }_{\mathrm{charge\ (\LL)}}
  \     \times  \ 
  \underbrace{ {\cal F}_{\JJ}^{N_{O_{\JJ}}}\,  {(1- {\cal F}_{\JJ} )}^{N_{S_{\JJ}}} }_{\mathrm{charge\ (\JJ)}} \ ,
  \label{eq:like3}
\end{equation}
\noindent where ${\cal F}_{\LL}$ is the expected fraction of lepton candidates
with charge opposite to the charge tag of the other hemisphere in sample $\LL$,
$N_{O_{\LL}}$ and $N_{S_{\LL}}$ are the numbers of candidates with opposite and same charge
found in the data. The same holds for sample $\JJ$.

The expected fractions ${\cal F}_{\LL}$ and  ${\cal F}_{\JJ}$ are sensitive to the relative
contributions of the $\bl$ and $\bcl$ components and
depend on the rate of correct tagging for the opposite hemisphere charge estimator,
as well as on the background components.

\boldmath
\subsection{Flavour composition of the selected samples}
\unboldmath
\label{sec:btageff}
The flavour 
composition for each of the three samples  is estimated as follows.
The fraction $\fhemi$ of single b-tagged hemispheres is measured from the data.
The efficiency $\epsilon_{\mathrm c}$  for tagging charm events
and the average efficiency  $\epsilon_{\mathrm{uds}}$ for tagging light quark events
are determined using simulated events, and the sample composition
is calculated as
\begin{eqnarray*}
  \fhemib & = & 1 - \frac{\Rc  \epsilon_{\mathrm c}  +
 (1- \Rb - \Rc) \epsilon_{\mathrm{uds}}} {\fhemi} \ , \\
  \fhemic & = & \frac{\Rc  \epsilon_{\mathrm c}}{\fhemi} \ , \\
  \fhemix & = & \frac{ (1- \Rb - \Rc) \, \epsilon_{\mathrm{uds}}}{\fhemi} \ ,
\end{eqnarray*}
where $\Rb$ and  $\Rc$~\cite{summer00}  are the ratios of the $\bb$ and $\cc$  
partial widths of the Z to the total hadronic width,
taken from experimental measurements.

\boldmath
\subsection{Charge tagging in samples $\LL$ and $\JJ$}
\unboldmath
\label{sec:chargetag}

The terms  ${\cal F}_{\LL}$ and 
 ${\cal F}_{\JJ}$ in Eq.~\ref{eq:like3} 
are written in terms of the probabilities that the charge
of the parent quark in the hemisphere opposite to the
lepton candidate is correctly tagged. 
These probabilities for  b or  $\bar{\mathrm {b}}$ quarks
$\Pb$, discussed in Section~\ref{sec:method},
are measured from the data by means
of a double-tag method. The description below applies to both
samples $\LL$ and $\JJ$.

The charge-tagging selection cut is applied to both hemispheres, and the
events selected are split into three classes: 
those in which the two hemispheres have
opposite charge, and those in which have the same charge, positive or negative.
The three fractions $\Foc$, $\Fpp$ and   $\Fmm$ can be written as
\begin{equation}
  \label{foc_data}
  \Foc \ = \  \fevtb \Focb + \fevtc \Focc + \fevtx \Focx \ ,
\end{equation}
\noindent  where  $\fevtb$, $\fevtc$ and $\fevtx$ are the contributions
of b, c and light flavour events to this sample,
with similar equations for $\Fpp$ and  $\Fmm$. 
The different flavour contributions are 
measured with the same procedure used to estimate the hemisphere sample
composition (Section~\ref{sec:btageff}): 
\begin{eqnarray}
  \nonumber
  \fevtb & = & 1 - \frac{\Rc  \zeta_{\mathrm c}  +
    (1- \Rb - \Rc) \zeta_{\mathrm{uds}}} {\fevt} \ , \\
  \nonumber
  \fevtc & = & \frac{\Rc  \zeta_{\mathrm c}}{\fevt} \ , \\
  \fevtx & = & \frac{ (1- \Rb - \Rc) \, \zeta_{\mathrm{uds}}}{\fevt} \ ,
\end{eqnarray}
where $\fevt$ is the fraction of events with both hemispheres tagged
in the data, and $\zeta_{\mathrm c}$ and
$\zeta_{\mathrm{uds}}$ are the charm and light quark event 
efficiencies determined from the simulation.

The fractions $\Focc$ ($\Fppc$, $\Fmmc$) and  $\Focx$ ($\Fppx$, $\Fmmx$)
of events with opposite (same) hemisphere charges in charm and light quark events 
are taken from the simulation,
and Eq.~\ref{foc_data} is solved for $\Focb$ ($\Fppb$,   $\Fmmb$). 

In order to extract $\Pb$ from $\Focb$, several effects must be considered.
\begin{itemize}
\item Because of the interaction of particles 
  with the detector material, the probability for correct
  charge tagging is different for b and $\bar {\mathrm {b}}$ quarks. 
  The difference can be defined as
  $\Pbm = \Pbdt (1-\epsilon)$, $\Pbp = \Pbdt (1+\epsilon)$, where  $\Pbdt$ is the average
  probability of correct charge tagging in the double-tagged sample selected.
\item The probabilities of correct charge  tagging in the two hemispheres
  are not independent. The correlation $\rho$ is defined so that 
  a hemisphere opposite to one in which the charge was correctly tagged
  has the probability of correct charge  tagging enhanced by a factor
  $(1+\rho)$.
\item The sample used for the branching ratio measurement 
  includes hemispheres for which
  the opposite one did not fulfil the selection cuts. The selection
  in the opposite hemisphere has some correlation with the 
  charge tagging probability of the hemisphere considered, therefore 
  the parameter which is measured with the double tag method 
  requires a further correction, $\alpha$, as explained below.
\end {itemize}

With all these effects taken into account, $\Pbdt$ is related to  $\Focb$ via
\begin{equation}
  \label{foc_b}
  \Focb \ = \ {\left( {\Pbdt}\right)}^2 + 
  {\left(1 - \Pbdt \right)}^2 +2  {\left( {\Pbdt}\right)}^2 
\left(\rho - \epsilon^2 - \epsilon^2 \rho\ \right) ,
\end{equation} 
where $\epsilon$ can be written as
\begin{equation}
  \label{fppmm_b}
   \epsilon = \frac{\Fppb - \Fmmb}{2 \Pbdt} .
\end{equation} 
Eq.~\ref{foc_b} and~\ref{fppmm_b} can be solved for $\Pbdt$ and $\epsilon$ 
taking for
$\rho$ the estimate provided by the simulation. Then $\Pb$ is calculated
as 
\begin{equation} 
  \label{last}
  \Pb = (1-\alpha) \Pbdt \ , 
\end{equation} 
\noindent in terms of a bias correction $\alpha$ also taken from
simulated events. Both $\alpha$ and $\rho$ are found to be smaller 
than 1\%.

\section{Sources of systematic errors}
\label{sec:syst}
In this section the sources of possible
systematic effects are discussed.
The estimated uncertainties are summarized  in Tables~\ref{sys_table}
and~\ref{sys_mod_table}.

\begin{table}[tb!]
\begin{center}
\begin{small}
\begin{tabular}{|l||c|c||c|c||}
 \hline
 \hline
\rule{0pt}{5mm}
   &  \multicolumn{2}{c||}{\normalsize $\Delta [ \BRbl ]$ }  & \multicolumn{2}{|c||}{\normalsize $\Delta [ \BRbcl ]$} \\
[0.11cm]
 \hline
  Source    \rule{0pt}{4.6mm}  & $\pt$    &   Charge    & $\pt$       & Charge       \\[0.09cm]
\hline
$\Rb$  \rule{0pt}{4mm}      & ~~~~~negl.~~~~~ & ~~~~~negl.~~~~~ & ~~~~~negl.~~~~~ & ~~~~~negl.~~~~~    \\
$\Rc$                       & $\pm$ 0.005 & $\mp$ 0.007 & $\mp$ 0.002 & $\pm$ 0.017  \\
$\gsbb$                     & $\mp$ 0.002 & $\mp$ 0.002 & $\mp$ 0.002 & $\mp$ 0.001  \\
$\gscc$                     & $\mp$ 0.001 & $\mp$ 0.006 & $\mp$ 0.014 & $\mp$ 0.006  \\
electron ID efficiency      & $\mp$ 0.063 & $\mp$ 0.081 & $\mp$ 0.087 & $\mp$ 0.056  \\
$\gamma$ conversions        & $\pm$ 0.003 & $\mp$ 0.006 & $\mp$ 0.022 & $\mp$ 0.008  \\ 
electron bkg                & $\pm$ 0.004 & $\mp$ 0.007 & $\mp$ 0.026 & $\mp$ 0.009  \\
muon ID     efficiency      & $\pm$ 0.065 & $\pm$ 0.063 & $\pm$ 0.039 & $\pm$ 0.039  \\
muon   bkg                  & $\pm$ 0.002 & $\mp$ 0.013 & $\mp$ 0.037 & $\mp$ 0.015  \\
$\BRbcll$                   & $\pm$ 0.004 & $\pm$ 0.022 & $\pm$ 0.002 & $\mp$ 0.026  \\
$\BRJpsill$                 &     negl.   &       negl. &     negl.   &    negl.     \\
$\BRbtaul$                  & $\mp$ 0.017 & $\mp$ 0.043 & $\mp$ 0.053 & $\mp$ 0.011  \\ 
BR($\bcbl$)                 & $\pm$ 0.010 & $\mp$ 0.223 & $\mp$ 0.407 & $\mp$ 0.039  \\ 
BR($\cl$)                   &  negl.      & $\mp$ 0.016 & $\mp$ 0.009 & $\pm$ 0.016  \\ 
BR($\bul$)                  & $\mp$ 0.032 & $\mp$ 0.022 & $\pm$ 0.013 & $\mp$ 0.004  \\ 
b fragmentation             & $\mp$ 0.074 & $\mp$ 0.089 & $\mp$ 0.120 & $\mp$ 0.101  \\
c fragmentation             & $\pm$ 0.001 & $\pm$ 0.005 &  negl.      & $\mp$ 0.005   \\
$\epsilon_{\mathrm    c}$ sample $\BB$    & $\pm$ 0.027 & $\pm$ 0.015 & $\mp$ 0.009 & $\mp$ 0.010  \\
$\epsilon_{\mathrm{uds}}$ sample $\BB$    & $\pm$ 0.015 & $\pm$ 0.016 & $\pm$ 0.012 & $\pm$ 0.011  \\
$\epsilon_{\mathrm    c}$ sample $\JJ$    & -           & $\mp$ 0.018 & -           & $\pm$ 0.029  \\
$\epsilon_{\mathrm{uds}}$ sample $\JJ$    & -           & negl.       & -           & negl.  \\
$\epsilon_{\mathrm    c}$ sample $\LL$    & -           & $\mp$ 0.012 & -           & $\pm$ 0.019  \\
$\epsilon_{\mathrm{uds}}$ sample $\LL$    & -           & negl.       & -           & negl.  \\
c charge tag rate           & -           & $\pm$ 0.036 & -           & $\mp$ 0.057   \\
b charge tag rate           & -           & $\pm$ 0.069 & -           & $\mp$ 0.109   \\
Mixing in $\bl$             & -           & $\pm$ 0.035 & -           & $\mp$ 0.055  \\
Mixing in $\bcl$            & -           & $\mp$ 0.055 & -           & $\pm$ 0.087  \\
bkg charge correlation      & -           & $\pm$ 0.027 & -           & $\mp$ 0.043  \\ 
b tag - lept correlation   & $\pm$ 0.006 & $\mp$ 0.007 & $\mp$ 0.025 & $\mp$ 0.005   \\[0.07cm]

\hline
\hline
 {\bf Total} \rule{0pt}{4.6mm}    & $\pm$ 0.128 & $\pm$ 0.290 & $\pm$ 0.443 & $\pm$ 0.212  \\[0.09cm]
\hline
\hline
\end{tabular}
\end{small}
\caption[.]{\small Estimated contributions to the systematic uncertainties on 
$\BRbl$ and \mbox{$\BRbcl$}). Results for both transverse momentum 
and charge correlation analyses are given. 
Uncertainties related to the modelling of semileptonic decays are shown separately in Table~\ref{sys_mod_table}.
All values are given in units of $10^{-2}$.}
\label{sys_table}
\end{center}
\end{table}

\begin{table}[tb!]
\begin{center}
\begin{small}
\begin{tabular}{|l||c|c||c|c||}
 \hline
 \hline
\rule{0pt}{5mm}
   &  \multicolumn{2}{c||}{\normalsize $\Delta [ \BRbl ]$ }  & \multicolumn{2}{|c||}{\normalsize $\Delta [ \BRbcl ]$} \\
[0.11cm]
 \hline
  Source    \rule{0pt}{4.6mm}  & $\pt$    &   Charge    & $\pt$       & Charge       \\[0.09cm]
\hline
${\mathrm D} \ell  \nu$  \rule{0pt}{4mm}   & ~~~$\pm$ 0.012~~~ & ~~~$\pm$ 0.022~~~  & ~~~$\pm$ 0.029~~~ & ~~~$\pm$ 0.013~~~  \\
$\Dstar \ell \nu$                          & $\mp$ 0.008 & $\pm$ 0.077  & $\pm$ 0.035 & $\pm$ 0.010  \\
Inclusive  $\Dand  {\mathrm {X}} \ell \nu$ & $\pm$ 0.254 & $\pm$ 0.086  & $\mp$ 0.265 & $\mp$ 0.004  \\
$\Done  \ell \nu$                          & $\mp$ 0.068 & $\mp$ 0.028  & $\pm$ 0.056 & $\mp$ 0.007  \\
$\Dtwo  \ell \nu$                          & $\mp$ 0.047 & $\mp$ 0.018  & $\pm$ 0.040 & $\mp$ 0.006  \\
Broad states                               & $\pm$ 0.327 & $\pm$ 0.169  & $\mp$ 0.191 & $\pm$ 0.052  \\[0.05cm]
\hline
$\Bs$ fraction      \rule{0pt}{4mm}        & $\pm$ 0.049 & $\pm$ 0.060  & $\pm$ 0.063 & $\pm$ 0.048  \\
Baryon  fraction                           & $\mp$ 0.025 & $\mp$ 0.012  & $\pm$ 0.059 & $\pm$ 0.044  \\[0.05cm]

\hline
\hline
$\bl$  modelling \rule{0pt}{5mm}& $\pm$ 0.426          & $\pm$ 0.202           & $\pm$ 0.348         & $\pm$ 0.085      \\[0.05cm]
$\cl$ modelling \rule{0pt}{5mm}& $^{-0.087}_{+0.072}$ & $^{-0.038}_{+0.021}$ & $^{-0.037}_{+0.020}$ & $^{-0.117}_{+0.063}$  \\[0.05cm]
$\bD$ modelling \rule{0pt}{5mm}& $^{-0.072}_{+0.060}$ & $^{-0.002}_{+0.001}$ & $^{-0.055}_{+0.049}$ & $^{+0.058}_{-0.050}$  \\[0.12cm]
\hline
\hline
{\bf Total modelling}  \rule{0pt}{5mm} & $^{+0.436}_{-0.441}$ & $^{+0.204}_{-0.206}$ & $^{+0.353}_{-0.355}$ & $^{+0.117}_{-0.156}$ \\
[0.12cm]
\hline
\hline
\end{tabular}
\end{small}
\caption[.]{\small Estimated contributions to the systematic uncertainties on 
$\BRbl$ and \mbox{$\BRbcl$}) related to the modelling
of semileptonic decays. The uncertainty in the modelling of direct $\bl$ decays
is estimated by adding in quadrature the uncertainties from the rate of the different
charmed mesons, inflated by 25\%, and the uncertainties from the production rates
of $\Bs$ mesons and b baryons, as discussed in Section~\ref{sec:newmodels}.
Results for both transverse momentum 
and charge correlation analyses are shown. 
All values are given in units of $10^{-2}$.}
\label{sys_mod_table}
\end{center}
\end{table}

\boldmath
\subsection{Z partial widths to $\bb$ and $\cc$}
\unboldmath
The values of $\Rb$ and $\Rc$ are used in the derivation of
the sample compositions of the three hemisphere samples, and
in the calculation of $\Pbl$ and $\Pbj$.
The most recent averages are taken~\cite{summer00},
$\Rb = 0.21653\pm 0.00069$
and $\Rc = 0.1709\pm 0.0034$, and the experimental errors
are considered as sources of systematic uncertainty.

\subsection{Heavy quarks from gluon splitting}
Charm and bottom quark pairs may be produced from the splitting of hard gluons.
The heavy flavour hadrons resulting from this process have a  significantly
softer energy spectrum and thus give rise to a source of 
prompt leptons with kinematic
properties substantially different from those produced by
heavy hadrons from direct Z decay. In addition,  leptons originating from
gluons splitting to heavy quarks have a random charge correlation
with the charge estimators defined in the opposite hemisphere.

The latest world average values are used for the number  of gluons
splitting to heavy quarks per hadronic Z decay~\cite{LEPreport}, 
\begin{eqnarray*}
\gbb & = & 0.00254 \pm 0.00051 \ , \\
\gcc & = & 0.0296\ \, \pm 0.038 \ ,
\end{eqnarray*}
\noindent and the experimental errors are used to estimate
the associated uncertainty.

\subsection{Electron identification efficiency and background}
\label{sec:elesys}

The electron identification efficiency is measured from data using 
photon conversions in the detector material. Correction factors are
derived, with respect to the Monte Carlo, for the dependence   
on momentum, transverse momentum and polar angle.
These factors typically differ from unity by less than 1\%.
The associated systematic uncertainty is estimated 
by removing the corrections.

The background from hadron misidentification is estimated by 
removing the cut on the energy loss, and studying the shape of the
$I$ estimator. This study is performed on data and Monte Carlo and 
no significant deviation is
observed. An uncertainty of 20\% on the backgroud level 
is assigned from the statistical
precision of the method.

The background from unidentified  photon conversions is estimated
by studying  the shape of the variable 
\[
\rho_\gamma \ = q\, d_0\, \ptbeam \ , 
\]
\noindent where $q$ is the charge of the electron, 
$d_0$ is the distance of closest
approach to the primary vertex in the $xy$ plane
signed according to the track angular momentum around the origin,  
and $\ptbeam$ is the component of the
track momentum transverse to the beam axis. The variable $\rho_\gamma$ is
expected to be zero for prompt electrons and proportional to the 
square of the materialization
radius for electrons coming from  photon conversions. The study of the 
positive tail of the distribution yields a correction factor of 1.05
to be applied to the simulation, with a statistical error of 0.02. 
The correction factor is removed to estimate the systematic uncertainty.

\subsection{Muon identification efficiency and background}
\label{sec:musys}
The identification efficiency for high energy muons is measured from the
data using Z decays to muon pairs, as a function of polar and azimuthal 
angle. Simulated events are reweighted to reproduce the measured efficiencies. 
Correction factors are typically a few per mil.

Simulated events show  some dependence of the identification
efficiency upon the muon momentum  for momenta around $3~\gevc$.
This effect is also checked on real data using 
$\gamma\gamma \to \mu^+ \mu^-$ events. 
Additional correction factors, of order 1 to  2\% are derived 
for muons with momenta between $2.5~\gevc$ and $4~\gevc$.
The shift observed when the corrections factors are removed is used
as an estimate of the systematic uncertainty.

The main background for muon candidates is due to misidentified pions as well as
pions decaying   before entering the calorimeters. The corresponding
contributions for kaons are substantially reduced by the cut on the measured
energy loss, and are estimated to be a factor of four smaller.

In order to check the background rate from data, $\KS\to \pi^+ \pi^-$
decays are selected in hadronic events to  
yield a 99\% pure sample of pions. The muon identification procedure 
is applied to these tracks, and the selection efficiency is compared between
data and Monte Carlo. Agreement is found within the statistical precision of the
test, which is 5\%. The test is repeated applying different $\btag$ cuts
in the hemisphere opposite to the $\KS$ candidate in order to check for
a possible dependence on the flavour. No trend is observed.

The uncertainty of 5\% estimated from the check with $\KS\to \pi^+ \pi^-$
is enlarged to 10\% for the assignment of a systematic error to the muon background.
This allows for additional uncertainties from the smaller kaon component,
as well as possible differences between data and simulation 
in the production rates and kinematic properties of pions and kaons in b events.

\boldmath
\subsection{Specific $\bcl$ correction}
\unboldmath
The $\bcl$ rate is in principle different
in transitions where the W from the b hadron decays leptonically, $\BRbcll$,  
or hadronically, $\BRbclh$.
The fit could yield a biased result for the average $\BRbcl$ if the acceptance
were  different for the two cases. This effect is investigated by changing in the 
simulation the relative population of the two species and recalculating all
efficiencies and spectra. The $\BRbcll$ is increased to a conservative 
20\% and the $\BRbclh$ is decreased accordingly. 
The shift observed  in the fitted values is taken
as an estimate of the associated systematic error.

\subsection{Other sources of prompt leptons}
\label{sec.othlep}

The rates of leptons coming from $\Jpsi$ and from 
intermediate $\tau$ decays used for this analysis are~\cite{LEPreport}

\begin{eqnarray*}
\BRJpsill &=& 0.00072 \ \, \pm 0.00006 \ , \\
\BRbtaul &=& 0.00419 \pm 0.00055 \ .
\end{eqnarray*}

Leptons produced from cascade b decays where the intermediate
charm is produced from a ${\mathrm W}\to  \bar{\mathrm c} {\mathrm s}$ transition,
denoted $\bcbl$, are also a background to the analysis.
They affect most directly the result for $\BRbcl$ in the transverse momentum analysis,
as they have kinematic properties similar to $\bcl$ decays. 
On the other hand
in the charge correlation analysis only the value of $\BRbl$ depends on the rate
of these transitions since the correlation between the charge of the lepton
and the charge of the parent quark is the same as in $\bl$ direct decays.
The value used is~\cite{LEPreport}
\[ \BRbcbl \ = \ 0.0162 \pm 0.0044 \ .
\]

The semileptonic decays  of charmed hadrons in 
the residual background of $\cc$ events contribute to the total
observed lepton yield.
The LEP average value~\cite{summer00} is taken:
\[
\BRcl \ = \ 0.0984 \pm 0.0032 \ .
\] 

\noindent All the above sources of leptons are subtracted from the total 
lepton yield.

The rate of lepton production in  charmless semileptonic b decays, 
\mbox{$\bul$},
mainly affects the high end of the $\pt$ spectrum. A variation
of this contribution  relative to the total BR($\bl$) 
is included in the systematic error calculation. 
The value used in this analysis is~\cite{roudeau}
\[
\BRbul \ = \ 0.00167 \pm 0.00055 \ .
\]

\subsection{Fragmentation of b quarks}
The scaled energy spectrum of b hadrons in the
simulation is modified in order to reproduce
the spectrum reconstructed in the model-independent
analysis of~\cite{tom}.

The statistical and systematic uncertainties on the population 
of each energy bin are propagated to the measured branching ratios,
taking into account bin-to-bin correlations. 
The systematic errors on the energy spectrum due to the 
uncertainty on the charmed meson species produced in B meson decays 
are not considered here, as they are correlated with the
uncertainty on the modelling of  semileptonic decays (Section~\ref{sec:bmod}).

The correction factors derived from the comparison of the measured and simulated
energy spectra of B mesons are also applied to $\Bs$ mesons and b baryons.

\subsection{Fragmentation of c quarks}
Charm fragmentation is simulated using the Peterson {\it et al.} model~\cite{peterson}.
The parameter $\varepsilon_{\mathrm c}$,  which 
controls the shape of the function, is adjusted to reproduce 
the measured value of the mean scaled energy 
of weakly-decaying charmed hadrons,  
\mbox{$\langle x_{\mathrm c}\rangle = 0.484 \pm 0.008$~\cite{LEPreport}},
and then varied by its uncertainty.

\subsection{Charm and light quark background}
The calculation of the sample compositions described in Section~\ref{sec:btageff}
relies on the simulation for estimates of the charm and light quark
hemisphere selection efficiencies. For samples $\BB$ and $\JJ$ the estimate of 
the uncertainty on the background efficiencies follows the lines of~\cite{ian},
where all the relevant physics parameters in the Monte Carlo have been
reweighted. 
The  values  used for this analysis are
\begin{eqnarray*}
 \epsilon_{\mathrm    c}^\BB \ = \ 0.00939      \pm 0.00094     \ \ \ & & 
 \epsilon_{\mathrm{uds}}^\BB \ = \ 0.00060      \pm 0.00015   \ ,   \\
 \epsilon_{\mathrm    c}^\JJ \ = \ 0.0439\ \,   \pm 0.0022\  \,  \ \ \ & & 
 \epsilon_{\mathrm{uds}}^\JJ \ = \ 0.0054\ \,   \pm 0.0008\  \,  \ . \\
\end{eqnarray*}

For  sample $\LL$ the dominant sources of uncertainty are the 
charm semileptonic branching fraction and decay modelling,
for charm hemispheres, and lepton background for light quark
hemispheres. The corresponding values and errors are 
\begin{eqnarray*}
 \epsilon_{\mathrm    c}^\LL \ = \ 0.0108  \, \pm 0.0011    \ \ \ & & 
 \epsilon_{\mathrm{uds}}^\LL \ = \ 0.0016  \, \pm 0.0003    \ .
\end{eqnarray*}

\subsection{Charge tagging in charm and light quark  hemispheres}

The fractions of opposite charge charm and light quark hemispheres 
entering Eq.~\ref{foc_data} are taken from the simulation. The 
difference between $\Fpp$ and $\Fmm$ for these flavours is neglected
in the analysis, therefore they are determined once
$\Foc$ is obtained.

In sample $\LL$,  $\Focc$ is determined by the relative amounts 
of true and fake leptons selected, and by the degree of 
correlation between the charge of fake leptons and the charge of
the c quark. The rates of lepton production from c hadron semileptonic decays
and from hadron misidentification are varied within their estimated
uncertainties (Sections~\ref{sec:elesys},~\ref{sec:musys} and~\ref{sec.othlep}).
The charge correlation for fake leptons is set to zero and half of the difference in $\Focc$ 
is taken as an estimate of the related uncertainties. 
Adding the uncertainties from the three sources in quadrature 
yields $\Focc = 0.87\pm0.04$ in sample  $\LL$,
corresponding to $\Pcl = 0.93 \pm 0.02$.

For sample $\JJ$ an estimate of the uncertainty on $\Focc$
is obtained by comparing the value with  $\Focb$.
For charm hemispheres, the correlation between the 
charge of the quark and the sign of the estimator is stronger
than in the case of b hemispheres. This is valid
separately for the two estimators $Q_p$ and $Q_s$. For the first,
the reason is  the higher charm quark charge. In the case
of  $Q_s$, the effect is due to the b-tagging cut
which selects charm hemispheres with particularly long lifetime,
and to the smaller charged particle multiplicity of c hadron
decays compared to b hadron decays.
The systematic uncertainty is evaluated by
setting the charge correlation for charm hemispheres equal
to that for b hemispheres and taking half of the 
difference as the systematic uncertainty; this procedure 
yields $\Focc = 0.64 \pm 0.02 $ in sample   $\JJ$,
corresponding to $\Pcj = 0.77 \pm 0.02$.

In light quark events, $\Focx$ is still somewhat 
larger than 0.5 due to the correlation of the
estimator with the parent quark charge, for both samples.
However the systematic uncertainty obtained by setting 
it equal to 0.5 is negligible.

\subsection{Charge tagging in b hemispheres}

Besides the sources already considered, the uncertainty 
on $\Pb$ also depends on the uncertainties on
$\Foc$, $\Fpp$, $\Fmm$,  $\fevt$, $\zeta_{\mathrm c}$,  $\zeta_{\mathrm{uds}}$,
$\rho$ and  $\alpha$
appearing in \mbox{Eq.~\ref{foc_data}--\ref{last}}.

The statistical errors on $\Foc$, $\Fpp$, $\Fmm$ and $\fevt$, 
which are measured from the data, are propagated to the
results of the fit and included in the statistical errors on the
branching ratios.

For sample $\JJ$ the values of $\rho$ and $\alpha$ measured in the simulated
events are 0.0096 and 0.0063, respectively.
For sample $\LL$,  $\alpha$ is found to be 0.0062, while there is no
significant correlation between the two hemispheres.
The associated systematic uncertainty is
estimated by setting  $\rho$ and $\alpha$ to zero and 
taking half of the  observed shift.

The uncertainties  from  $\zeta_{\mathrm c}$ and  $\zeta_{\mathrm{uds}}$
are negligible.

\subsection{Neutral B meson mixing}
The mixing of neutral B mesons contributes to the degradation of
the correlation between the charge of the lepton and the
charge of the parent b quark produced in the Z decay. 
The LEP average $\chibar = 0.1194 \pm 0.0043$~\cite{summer00}  
is used as input in the likelihood. This value
is interpreted as the average mixing rate for the
b hadron mixture from $\bl$ decays.

The relative population of $\Bd$ and B$^+$ mesons
is not equal in \mbox{$\bcl$} decays, due to the different 
semileptonic branching ratios of D$^+$ and D$^0$ mesons,
leading to an effectively higher value of the average mixing parameter
\mbox{$\chibar^\prime = \chibar\, (1+\delta)$}. The value
\mbox{$\delta = 0.13$} is estimated from the simulation 
and is varied by 50\% to estimate the
systematic uncertainty.

As a consequence the expected fraction of events with hemispheres 
of opposite sign for the $\bl$ and $\bcl$ components is
\begin{eqnarray*}
\bl  \! :  & & {\mathcal F} \ = \ \Pb \, (1-\chibar ) + (1-\Pb) \,  \chibar \ ,  \\
\bcl \! : & & {\mathcal F} \ = \ \Pb  \, \chibar^\prime + (1-\Pb)  \, (1-\chibar^\prime) \ .
\end{eqnarray*}

\subsection{Charge correlation for lepton background}
Leptons coming from kaon and pion decays in flight
as well as misidentified kaons and pions  
retain some information about the 
charge of the primary quark, both in charm and in b events.
This effect must be taken into account when evaluating
the opposite-charge and same-charge fractions in the charge
correlation analysis.
The rate at which the information about the quark charge
is retained is measured in the Monte Carlo.
The systematic uncertainty is evaluated by taking half of the difference
between the Monte Carlo value and no charge correlation.

\subsection{Hemisphere-hemisphere correlation}
\label{sec.hcorr}

The b-tagging efficiency is affected by the presence
of a lepton in  the hemisphere opposite to that
used for the flavour tagging, in a way that depends on both
the momentum and the transverse momentum of the lepton.
In particular, hemispheres opposite to a high momentum
lepton are more likely to satisfy the b-tagging cut
since such leptons are likely to be produced in events
with little gluon emission, where both b hadrons have 
high momentum.
The b-tagging efficiency for those hemispheres 
is then higher than  expected in an unbiased sample of hemispheres.

The ratio $R = \epsilon^{l}_{\mathrm b}/(\epsilon_{\mathrm b}\times\epsilon_{l})$
is determined from simulated events,
where $\epsilon_{\mathrm b}$ and $\epsilon_{l}$ are respectively 
the b-tagging efficiency and the probability 
of finding a lepton in a b hemisphere while $\epsilon^{l}_{\mathrm b}$ is the 
probability that in the same event one hemisphere is b-tagged
and in the opposite hemisphere one lepton is selected. 
The dependence of $R$ upon the lepton momentum is irrelevant for the analysis,
while the overall value affects the part of the likelihood labelled
as ``counting'' in Eq.~\ref{eq:like2} and~\ref{eq:like3}, and the dependence
on the lepton transverse momentum affects the part labelled as ``$\pt \ {\mathrm{spectrum}}$'' 
in Eq.~\ref{eq:like2}. The quantity $R$ is therefore  determined as a function of the 
lepton $\pt$, for the different sources of lepton
candidates, and the corresponding correction factors are applied in the analysis.
Deviations from unity are typically smaller than 1\%. The 
corresponding systematic uncertainty is estimated by removing the corrections.

\subsection{Modelling of direct semileptonic b decays}
\label{sec:bmod}
\label{sec:newmodels}   
In the past, in all measurements of the inclusive $\bl$ rate 
the decay kinematics were modelled using the ACCMM spectrum,
and the associated systematic uncertainty was estimated
from the shift observed using the ISGW and ISGW$^{\star\star}$ spectra,
as proposed in~\cite{lephf_old}.

In this work a different approach is used, based on the available
measurements of the B meson decay rates into the different
charm hadron final states.
B$^{0}$ and B$^{+}$ mesons decay semileptonically
into D, $\Dstar$ and $\Dstst$ mesons, as well as non-resonant $\Dand \pi$ final states. 
Leptons coming from each of these
components have a different energy spectrum so that
the shape of the inclusive ${\mathrm B}\to {\mathrm {X}} \ell \nu$
spectrum depends on the branching fractions of  B$^{0}$ and B$^{+}$ into
the various charmed species. 

The fractions of D, $\Dstar$, $\Done$, and $\Dtwo$ in the simulation
are reweighted using the
latest measured values~\cite{roudeau,pdg2000} reported in Table~\ref{br_input}.
The broad $\Dstst$ states are assumed to be equal to the sum 
of the narrow $\Done$ and $\Dtwo$ states, and the non-resonant
$\Dand \pi$ decays account for the rate needed
in order to add up to the measured inclusive  $\Dand {\mathrm {X}}$ 
branching ratio.
  
\begin{table}[h!]
  \begin{center}
    \begin{tabular}{|l||l|}
      \hline
      Process                      & BR (\%) \\
      \hline
      \hline
      ${\mathrm B}\rightarrow {\mathrm D} \ell  \nu$        & $1.95    \pm 0.27$  \\
      ${\mathrm B}\rightarrow \Dstar \ell \nu$              & $5.05    \pm 0.25$  \\
      ${\mathrm B}\rightarrow \Dand  {\mathrm {X}} \ell \nu$& $2.7\ \, \pm 0.7$   \\
       with \ \   ${\mathrm B}\rightarrow \Done \ell \nu$  & $0.63    \pm 0.11$  \\
       with  \ \  ${\mathrm B}\rightarrow \Dtwo \ell \nu$  & $0.23    \pm 0.09$  \\
      \hline
    \end{tabular}
    \caption[.]{\small Branching ratios for semileptonic decays of B mesons 
      with different charmed mesons in the final state.}
    \label{br_input}
  \end{center}
\end{table}

The systematic uncertainty is evaluated by varying the measured 
branching fractions within their estimated errors. Additionally
the rate of the broad states is set to zero and compensated entirely
with the non-resonant D$^{(\star)}\pi$ states.
In each case, the B meson energy spectrum measured in~\cite{tom}
is used, thus taking into account the correlation with the
present analysis. This approach renders the analysis presented
independent of the modelling of the b fragmentation 
in the simulation.

\begin{figure}[bt!]
\begin{center}
 \mbox{\epsfig{figure=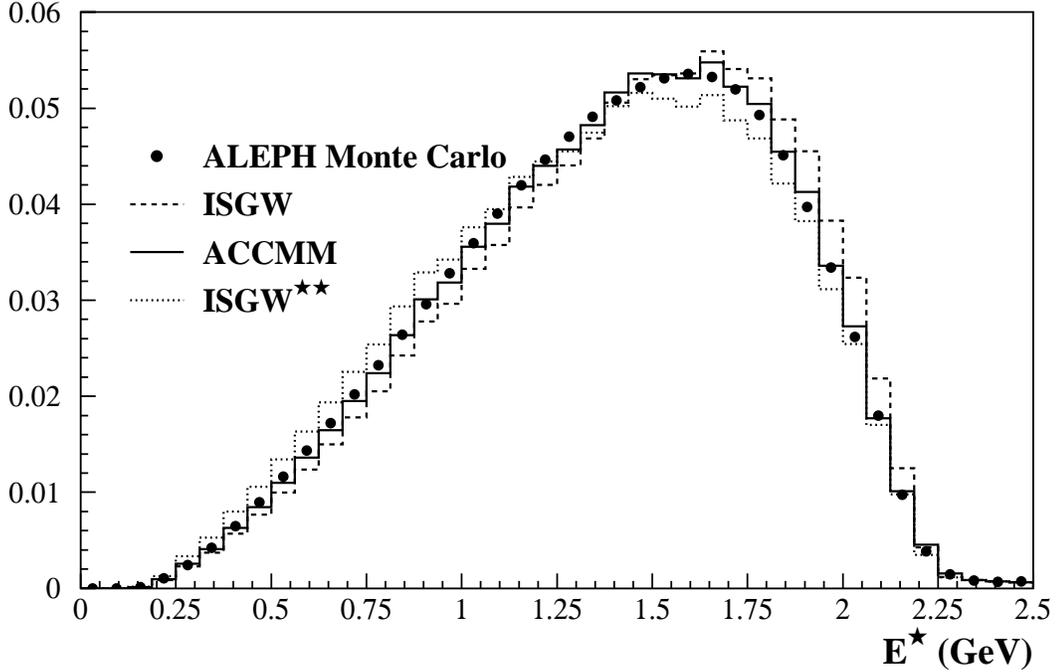,width=160mm}} 
     \vspace{-0.6cm}
\caption{\small Lepton energy spectrum in the b hadron rest frame. 
The histograms show the distributions
obtained after reweighting with the ACCMM, ISGW and ISGW$^{\star \star}$  models. The dots show
the spectrum after correcting the  Monte Carlo by following the procedure 
described in Section~\ref{sec:bmod} using the central value of the 
branching ratios listed in Table~\ref{br_input}.} 
\label{d_dstar_spectra}
\end{center}
\end{figure}

\begin{figure}[tb!]
  \begin{center}
    \mbox{\epsfig{figure=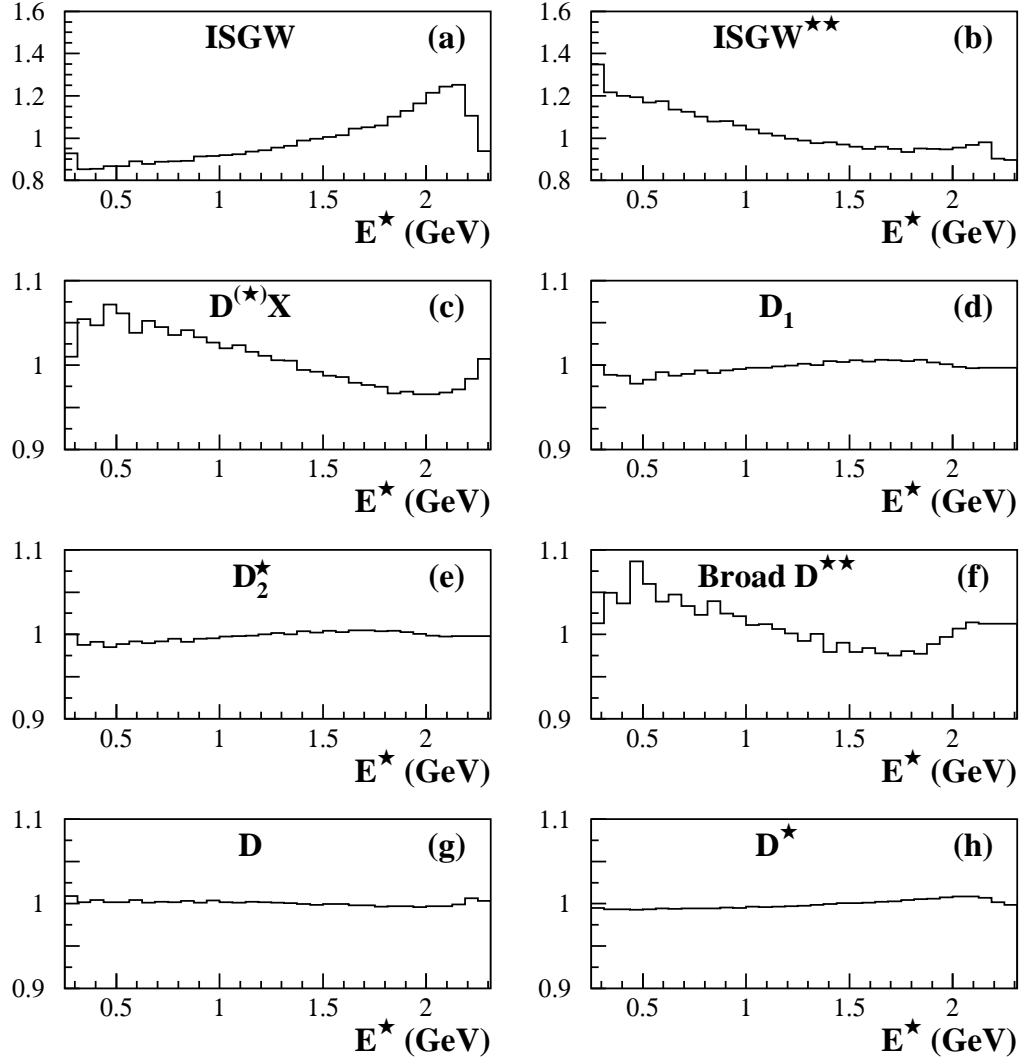,width=160mm}} 
     \vspace{-1cm}
    \caption{\small 
      The plots (a) and (b) show the ratio between the lepton spectra given by the 
      ISGW and ISGW$^{\star\star}$  models and
      the spectrum obtained by correcting the Monte Carlo using the central value of the
      branching ratios listed in Table~\ref{br_input}.
      Histograms (c)--(h) show the effect on the spectrum due to the variation  
      by one standard deviation of each component with respect to the central value.}
    \label{d_dstar_vari}
  \end{center}
\end{figure}

The energy spectra obtained with this procedure for 
the mixture of electrons  and muons used in the analysis
are compared in 
Fig.~\ref{d_dstar_spectra} with the inclusive spectra given in~\cite{lephf_old}.
In Fig.~\ref{d_dstar_vari} the effect of the corrections applied
to estimate the systematic uncertainty is shown.
The spectra resulting from this semi-exclusive treatment always
lie between the softest (ISGW$^{\star \star}$) and hardest (ISGW)
spectra.

The procedure described applies to B$^0$ and B$^+$ decays, which
represent about 80\% of the sample. Uncertainties in the production
rates of $\Bs$ mesons and b baryons can affect the analysis to 
the extent that their decay kinematics differ from those
of B$^0$ and  B$^+$ mesons. The $\Bs$ and b baryon production
rates are reweighted to the latest experimental results, f$_{\Bs} = 0.094 \pm 0.022$
and f$_{\mathrm {b-baryon}} = 0.101 \pm 0.018$~\cite{roudeau}. 
Uncertainties in the modelling of $\Bs$ and b baryon decays are
accounted for by enlarging the uncertainties from the branching
ratios of Table~\ref{br_input} by 25\%.
Finally, the observed shifts on $\BRbl$ and $\BRbcl$ are 
symmetrized and added in quadrature.

\boldmath
\subsection{Modelling of $\cl$ and $\bcl$ decays}
\unboldmath
The $\cl$ decay spectrum is obtained by means of  
a combined fit to DELCO~\cite{delco} and MARK III~\cite{marc3}
data, performed using the ACCMM model and varied as described in~\cite{lephf_old}. 

In order to describe $\bcl$ decays, the  model
proposed for $\cl$ is combined with the measured 
$\bD$ spectrum from CLEO~\cite{cleo_bd} following
the procedure of~\cite{lephf_old}.

\section{Results}

\begin{figure}[tb!]
\begin{center}
 \mbox{\epsfig{figure=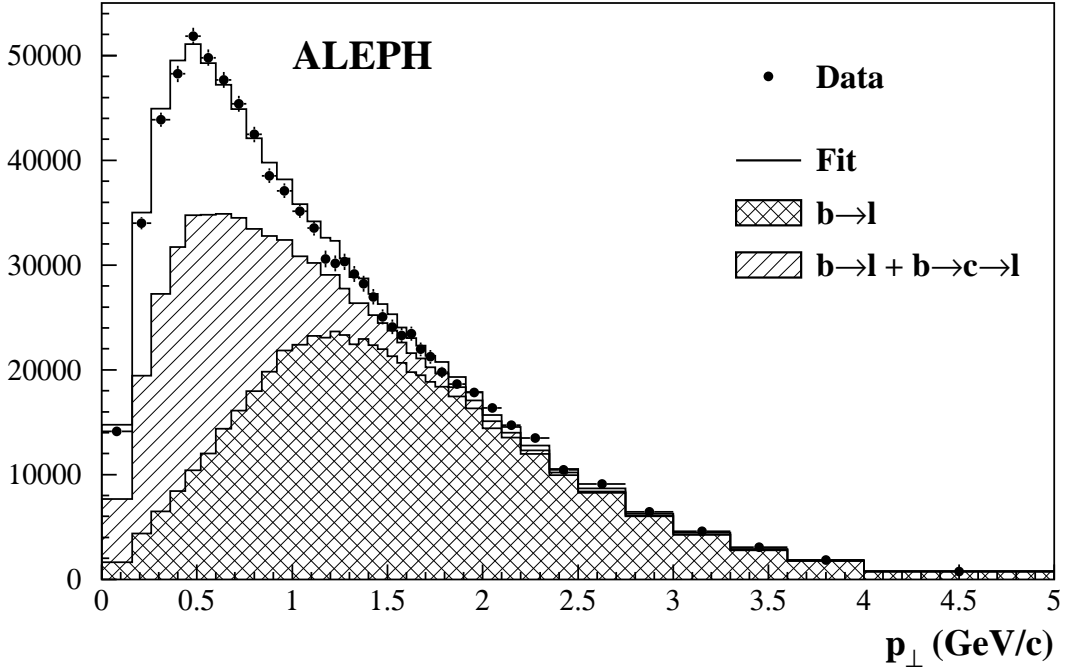,width=160mm}} 
\vspace{-0.8cm}
\caption{\small Comparison between the lepton $\pt$ spectrum measured in the data and the result of the fit.
The contribution from $\bl$ as well as the sum of $\bl$ and $\bcl$ are shown.
The error bars show statistical errors on the data.}
\label{fit_fig}
\end{center}
\end{figure}

The results obtained with the transverse momentum analysis  
are 
\vspace{-0.1cm}
      \begin{eqnarray}
        \BRbl             & = & 0.1107 \, 
        \pm 0.0007 \, _{\mathrm{stat}}\
        \pm 0.0013 \, _{\mathrm{syst}}\
        \pm 0.0044 \, _{\mathrm{model}}\ ,
        \nonumber \\
        \BRbcl            & = & 0.0752 \,
        \pm 0.0010 \, _{\mathrm{stat}} \
        \pm 0.0044 \, _{\mathrm{syst}}\
        \pm 0.0035 \, _{\mathrm{model}}\ \,  ,
        \nonumber 
      \end{eqnarray}

\noindent with a statistical correlation of $-0.45$.  
The fitted spectrum is compared to the $\pt$ distribution observed in the data in 
Fig.~\ref{fit_fig}, where the contributions of the two signal processes are also shown.

\begin{figure}[tb!]
\begin{center}
 \mbox{\epsfig{figure=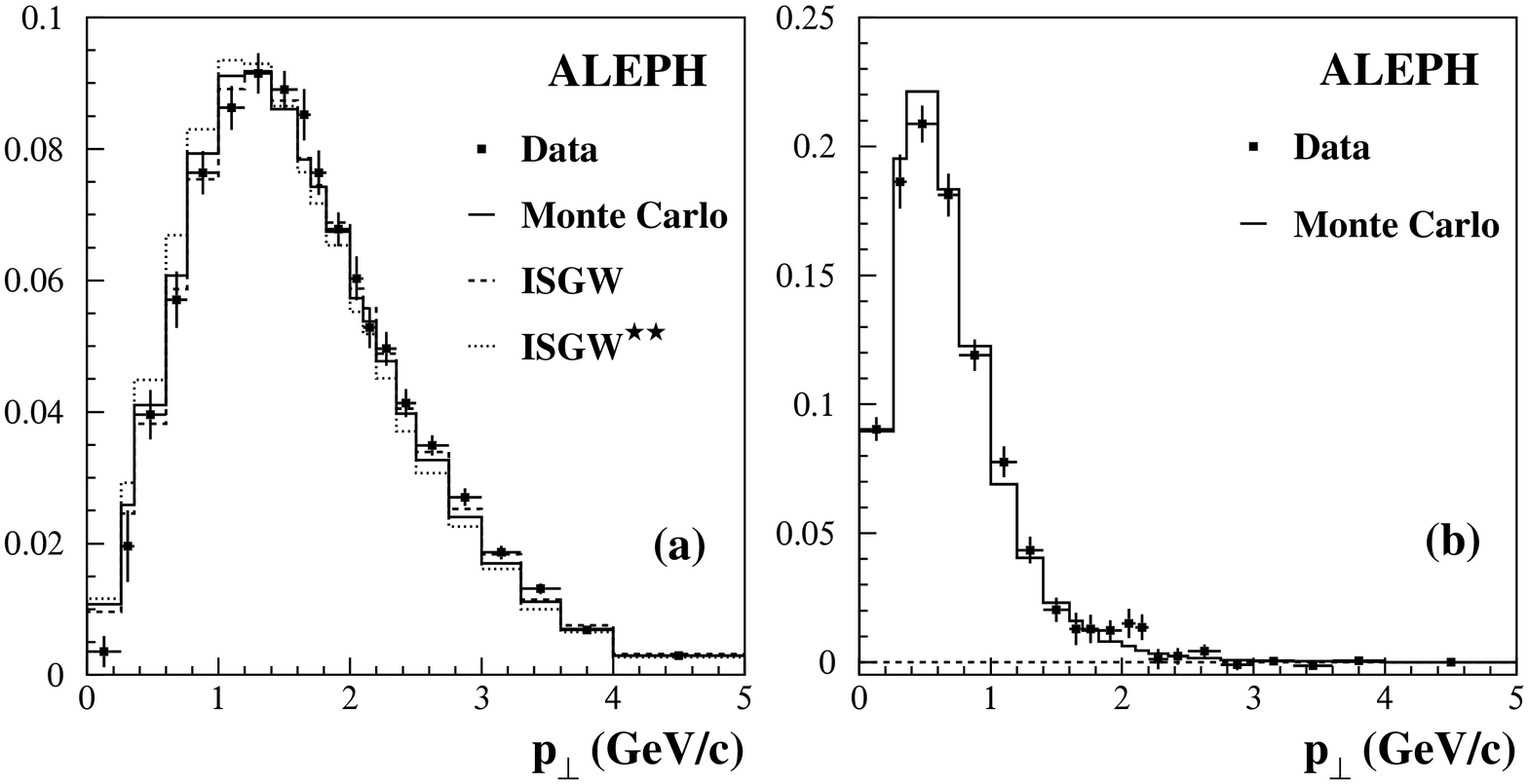,width=170mm}} 
\vspace{-1cm}
\caption{\small Comparison between the lepton $\pt$ spectra
measured in the data with the charge correlation 
analysis and that expected from simulation. 
(a)  $\bl$; data are compared with the simulation corrected as 
described in Section~\ref{sec:newmodels}, ISGW and ISGW$^{\star\star}$ 
models. (b) $\bcl$.
All the spectra are normalized to unit area.}
\label{pt_from_data_bl_bcl}
\end{center}
\end{figure}

The charge correlation analysis gives
\vspace{-0.8cm}

      \begin{eqnarray}
        \BRbl             & = & 0.1057 \, 
        \pm 0.0011 \, _{\mathrm{stat}}\
        \pm 0.0029 \, _{\mathrm{syst}}\
        \pm 0.0020 \, _{\mathrm{model}}\ ,
        \nonumber \\
        \BRbcl            & = & 0.0830 \,
        \pm 0.0016 \, _{\mathrm{stat}} \
        \pm 0.0021 \, _{\mathrm{syst}}\
        { ^{+0.0012}  _{-0.0016}}\, _{\mathrm{model}} \ ,
        \nonumber 
      \end{eqnarray}

\noindent with a statistical correlation of $-0.76$.
Since this method does not  use the information
from the lepton $\pt$, the $\pt$ spectrum can be extracted
from the data
by repeating the  charge correlation analysis in each $\pt$ bin. 
The values of \mbox{$\BRbl$} and \mbox{$\BRbcl$} measured in each $\pt$ bin
are then used to reweight the Monte Carlo spectrum previously
corrected as described in Section~\ref{sec:newmodels}.
In Fig.~\ref{pt_from_data_bl_bcl} 
the $\bl$ and $\bcl$ spectra shapes 
from the data are compared with the Monte Carlo.
The discrepancy observed reflects the difference between
the results obtained with the two methods.

The differences can be quantified taking into account
the correlation between systematic errors, as
$\Delta_{\bl}$ = 0.0049 $\pm$ 0.0036 (total error, of which 0.0025 is due to
modelling) and 
$\Delta_{\bcl}$ = 0.0077 $\pm$ 0.0056 (total error, of which 0.0039 is due 
to modelling).
The two results are therefore consistent within 1.4 times the total uncorrelated error.

The statistical errors are correlated and smaller than the systematic errors.
In the calculation of the average, maximal statistical correlation is assumed between the two
measurements, i.e. 81\% for $\BRbl$ and 85\% for \mbox{$\BRbcl$}, 
which leads to a slight overestimate of the statistical errors.
The weights used to calculate the average are
\mbox{W$_{\bl}(\pt)$= 0.25}, \mbox{W$_{\bl}$(charge)= 0.75}, 
\mbox{W$_{\bcl}(\pt)$= 0.15}, \mbox{W$_{\bcl}$(charge)= 0.85}. 
They are determined by  minimizing the total error (B.L.U.E. technique~\cite{blue}).
The resulting average values are
\vspace{-0.5cm}

      \begin{eqnarray}
        \BRbl         & = & 0.1070 \, 
        \pm 0.0010 \, _{\mathrm{stat}}\
        \pm 0.0023 \, _{\mathrm{syst}}\
        \pm 0.0026 \, _{\mathrm{model}}\ ,
        \nonumber \\
        \BRbcl        & = & 0.0818 \,
        \pm 0.0015 \, _{\mathrm{stat}} \
        \pm 0.0022 \, _{\mathrm{syst}}\
        { ^{+0.0010}  _{-0.0014}}\, _{\mathrm{model}} \ .
        \nonumber 
      \end{eqnarray}

\subsection{Consistency checks}

The analysis was performed seperately
for each year of data taking, to check for possible
detector-related effects. The results are consistent 
throughout the whole LEP I period.
The analysis was also performed using electron and muon 
candidates separately. The two species of leptons have 
different uncertainties in the selection efficiency
and in the rate of background. The tighter  selection
cut applied on the muon momentum is reflected in a different
dependence on modelling.
The results are consistent within errors as shown
by their differences quoted below.

\vspace{0.2cm}

\noindent{\bf Transverse momentum analysis:}
      \begin{eqnarray}
        \Delta^{{\mathrm {e}}-\mu}_{\bl}  & = & -0.0029   \,
        \pm 0.0013 \, _{\mathrm{stat}} \
        \pm 0.0018 \, _{\mathrm{syst}}\
        \pm 0.0002 \, _{\mathrm{model}}\ \,  ,
        \nonumber \\ 
        \Delta^{{\mathrm {e}}-\mu}_{\bcl}  & = & \ \  \, 0.0017   \,
        \pm 0.0020 \, _{\mathrm{stat}} \
        \pm 0.0021 \, _{\mathrm{syst}}\
        \pm 0.0011 \, _{\mathrm{model}}\ \,  .
        \nonumber 
      \end{eqnarray}

\noindent{\bf Charge correlation analysis:}
      \begin{eqnarray}
        \Delta^{{\mathrm {e}}-\mu}_{\bl}  & = & -0.0015   \,
        \pm 0.0021 \, _{\mathrm{stat}} \
        \pm 0.0021 \, _{\mathrm{syst}}\
        \pm 0.0004 \, _{\mathrm{model}}\ \,  ,
        \nonumber \\ 
        \Delta^{{\mathrm {e}}-\mu}_{\bcl}  & = & -0.0009   \,
        \pm 0.0031 \, _{\mathrm{stat}} \
        \pm 0.0017 \, _{\mathrm{syst}}\
        \pm 0.0005 \, _{\mathrm{model}}\ \,  .
        \nonumber 
      \end{eqnarray}

The cuts applied on the b-tagging variable, on the jet-charge value
and on the $\pt$ were also varied. The results, in both methods,
were found to be stable.

\section{Conclusions}

Two different strategies have been adopted to 
distinguish  the $\bl$ and the \mbox{$\bcl$} decays
in order to measure the inclusive semileptonic branching fractions of b hadrons,
$\BRbl$ and $\BRbcl$.
One measurement,  based on the analysis of the lepton
transverse momentum,  is significantly affected by the 
uncertainty from the modelling of the $\bl$ $\pt$ spectrum.  
The other is  based on the charge correlation between the lepton
and a charge estimator in the opposite hemisphere.
This has smaller dependence on the modelling of the 
decay kinematics, but the determination of the $\bl$ branching
fraction suffers from the uncertainty on the rate of 
$\bcbl$ decays. In both cases the modelling of the inclusive
$\bl$ decay kinematics is based on the measured rates of the 
different c hadrons in the final state.

The values of $\BRbl$ and  $\BRbcl$ measured with the two methods
are consistent  within  errors and they have been averaged. 
The resulting values are
\vspace{-0.2cm}
      \begin{eqnarray}
        \BRbl         & = & 0.1070 \, 
        \pm 0.0010 \, _{\mathrm{stat}}\
        \pm 0.0023 \, _{\mathrm{syst}}\
        \pm 0.0026 \, _{\mathrm{model}}\ ,
        \nonumber \\
        \BRbcl        & = & 0.0818 \,
        \pm 0.0015 \, _{\mathrm{stat}} \
        \pm 0.0022 \, _{\mathrm{syst}}\
        { ^{+0.0010}  _{-0.0014}}\, _{\mathrm{model}} \ ,
        \nonumber 
      \end{eqnarray}

\noindent in agreement with the latest measurements 
performed by other LEP experiments~\cite{bl_opal,bl_L3_1,bl_L3_2,bl_delphi},
and with the latest values from experiments running at the $\ups$
resonance~\cite{pdg2000}.

\section*{Acknowledgements}
We wish to thank our colleagues from the accelerator divisions
for the successful operation of LEP. It is also a pleasure to
thank the technical personnel of the collaborating institutions
for their support in constructing and maintaining the ALEPH 
experiment. Those of us  from non-member states
thank CERN for its hospitality.


\end{sloppypar}

\end{document}